\documentclass{acmplus}
\begin{document}

\def\viewstmt{\protect\raggedright
(Note: The HTML version of this paper is best
viewed using Microsoft Internet Explorer.
To view the HTML version using Netscape, add the following line to
your \texttt{\~{}/.Xdefaults} or \texttt{\~{}/.Xresources} file:
\par\noindent
\texttt{Netscape*documentFonts.charset*adobe-fontspecific:~iso-8859-1}
\par\noindent
For printing use the PDF version, as browsers may not print the
mathematics properly.)}

\long\def\howtoviewpdf{\thanks{\viewstmt}}
\long\def\howtoviewhtml{}

\newif\iftth
\tthfalse
\def\tthfig#1{\mbox{#1}}
\def\gap{\hspace{.6em}}

\title{The Structure of Broad Topics on the Web}
\author{%
\begin{tabular}{ccccccc}
Soumen Chakrabarti\thanks{Contact author, 
email \protect\path{soumen@cse.iitb.ac.in}} & \gap &
Mukul M. Joshi & \gap &
Kunal Punera & \gap &
David M. Pennock \\
IIT Bombay & &
IIT Bombay & &
IIT Bombay & &
NEC Research Institute
\end{tabular}  }
\maketitle

\begin{abstract}
The Web graph is a giant social network whose properties have been
measured and modeled extensively in recent years.  Most such studies
concentrate on the graph structure alone, and do not consider textual
properties of the nodes.  Consequently, Web communities have been
characterized purely in terms of graph structure and not on page
content.  We propose that a topic taxonomy such as Yahoo! or the Open
Directory provides a useful framework for understanding the structure
of content-based clusters and communities.  In particular, using a
topic taxonomy and an automatic classifier, we can measure the
background distribution of broad topics on the Web, and analyze the
capability of recent random walk algorithms to draw samples which
follow such distributions.  In addition, we can measure the
probability that a page about one broad topic will link to another
broad topic.  Extending this experiment, we can measure how quickly
topic context is lost while walking randomly on the Web graph.
Estimates of this topic mixing distance may explain why a global
PageRank is still meaningful in the context of broad queries.  In
general, our measurements may prove valuable in the design of
community-specific crawlers and link-based ranking systems.

\begingroup \raggedright
\par\smallskip
\paragraph{Categories and subject descriptors:}
H.5.4~[\textbf{Information interfaces and presentation}]:
Hypertext/hypermedia; \\
H.5.3~[\textbf{Information interfaces and presentation}]:
Group and Organization Interfaces, Theory and models; \\
H.1.0~[\textbf{Information systems}]: Models and principles.

\paragraph{General terms:}
Measurements, experimentation.

\endgroup

\par\smallskip \howtoviewhtml
\end{abstract}

\section{Introduction}

The Web is an actively evolving social network which brings together
myriad topics into a uniform hypertextual medium.  Superficially, it
looks like a giant graph where nodes are hypertext pages and edges are
hyperlinks.  A closer look reveals fascinating detail: the textual
matter, tag structure, site identity and organization, and (link)
affinity to prominent topic directories such as Yahoo! and the Open
Directory (DMoz),  to name just a few.  The best strategies for
browsing, searching and foraging for Web information are clearly
predicated on our understanding of the social processes which shape
the Web.  Naturally, the structure and evolution of the Web has been
under intensive measurements and modeling in recent years.

\subsection{Graph-theoretic models and measurements}

With a few notable exceptions, most studies conducted on the Web have
focused on its graph-theoretic aspects.


Barab\'asi and Albert \cite{Barabasi99c} proposed a local model
for social network evolution based on \emph{preferential attachment}:
nodes with large degree are proportionately more likely to become
incident to new links.  They applied it to the Web graph, and the
model predicted the power-law degree distribution quite accurately,
except for underestimating the density of low-degree nodes.  This
discrepancy was later removed by Pennock et 
al.~\cite{PennockLFLG2001winner} by using a linear combination of
preferential and random attachment.

Random graph models materialize edges independently with a fixed
probability.  If the Web were a random graph, large densely connected
components and bipartite cores would be extremely unlikely.  The Web
is not random, and such subgraphs abound on the Web.  A small
bipartite core is often an indicator of an emerging topic.  Kumar et
al.~\cite{RaviKumarRRT1999trawling} mine tens of thousands of such
bipartite cores and empirically observed that a large fraction are in
fact topically coherent, but the definition of a `topic' was left
informal.

Dense bipartite subgraphs are an important indicator of community
formation, but they may not be the only one.  Flake et 
al.~\cite{FlakeLLC2002selforg} propose a
general definition of a community as a subgraph whose internal link
density exceeds the density of connection to nodes outside it by some
margin.  They use this definition to drive
a crawler, starting from exemplary members of a community, and verify
that a coherent community graph is collected.

Br\"oder et al.~\cite{BroderKMRRSTW2000bowtie} exposed the large-scale
structure of the Web graph as having a central, strongly connected
core (SCC); a subgraph with directed paths leading into the SCC, a
component leading away from the SCC, and relatively isolated tendrils
attached to one of the three large subgraphs.  These four regions were
each about a fourth the size of the Web, which led the authors to call
this the ``bow-tie'' model of the Web.  They also measured interesting
properties like the average path lengths between connected nodes and
the distribution of in- and out-degree.  Follow-up work by Dill et 
al.~\cite{DillRMRST2001fractal} showed that subgraphs selected from the
Web as per specific criteria (domain restriction, occurrence of
keyword, etc.) also appear to often be bow-tie-like, although the
ratio of component sizes vary somewhat.  Content-based criteria used
keywords, not topics, and the interaction between topics, or the 
radius of topical clusters, were not addressed.

\subsection{Content-based locality measurements}

A few studies have concentrated on textual content.  

Davison pioneered a study \cite{Davison2000locality} over about 100,000
pages sampled from the repository of a research search engine called
\textsc{DiscoWeb}.  He collected the following kinds of pairs of Web
pages:
\begin{description}
\item[Random:]  
Two different pages were sampled uniformly at random (uar) from the
collection.
\item[Sibling:]
One page $x$ was picked uar from the collection, and then two distinct
random outlinks $u$ and $v$ were selected from among the outlinks
of~$x$.
\item[SameDomain:]
A page $u$ was sampled uar from the collection.  A random outlink
$v$ was chosen such that $u$ and $v$ were from the same host
(identified by name).
\item[DiffDomain:]
A page $u$ was sampled uar from the collection.  A random outlink
$v$ was chosen such that $u$ and $v$ were from \emph{different} hosts
(identified by name).
\end{description}
Davison represented the text on these pages using the standard
``vector space model'' from Information Retrieval \cite{SaltonM1983ir}
in which each document $u$ is represented by
a vector $\mathbf{u}$ of suitably normalized term counts
in a geometric space with an axis for each term.  He measured the dot
product between each pair of document vectors as a measure of
similarity, and made the following observations.
Random page pairs have almost nothing in common.  Linked pages are
more similar when the pages are from the same domain.  Sibling pages
are more similar than the linked pages of different domain.  More
recently, Menczer \cite{Menczer2001locality} has studied and modeled
carefully the rapid decay in TFIDF similarity to a starting node as
one walks away from that node.


A single starting page $u_0$ may be a noisy indicator of semantic
similarity with pages later on in the walk, because it may have a
limited vocabulary.  The pages on a short walk
$(u_0,u_1,\ldots,u_i)$  may
all be clearly about a given broad topic $c$, but each page may use a
negligible fraction of the vocabulary of $c$.  Thus, it is possible
that $\mathbf{u_0}\cdot\mathbf{u_i}$ is small, whereas
$\sum_c\Pr(c)\Pr(u_0|c)\Pr(u_i|c)$ (the probability that they are
both relevant to the same topic, see \S\ref{sec-intro-classify})
is quite large.

It is also more informative to estimate the distance at which
\emph{topical} memory is lost, rather
than only measure the rate of relevance decay for some fixed starting
topics, because there is no reference for how dissimilar $u_0$ and
$u_i$ need to be before we agree that memory of $u_0$ has been lost.

We also go beyond the studies of Davison and Menczer in that we
analyze the relation between different topics rather than
only model similarity between pages in a small neighborhood.

\subsection{Our contributions}

We bring together two existing experimental techniques to launch a
thorough study of topic-based properties of the Web: the ability to
\emph{classify} a Web page into predefined topics using a high-speed
automatic classifier, and the ability to draw near-uniform
\emph{samples} from the Web graph using random walks.  
We use a 482-class topic taxonomy from DMoz (\url{http://dmoz.org/})
and a sampling and classifying technique that we will describe
in \S\ref{sec-algo}.  By obtaining
evidence that our samples are faithful, we avoid processing large Web
crawls, although even our sampling experiments have fetched
almost 16 million pages.

Our study reveals the following properties about communities of broad
topics in the Web graph.

\paragraph{Convergence of topic distribution 
on undirected random walks:}
Algorithms for sampling Web pages uar have been evaluated on
structural properties such as degree distributions
\cite{BarYossefBCFW2000sample,RusmevichientongPLL2001sample}.
Extending these techniques,
we design a certain undirected random walk (i.e., assuming hyperlinks
are bidirectional) to estimate the distribution of Web pages w.r.t.\
the Dmoz topics (\S\ref{sec-base}).
We start from drastically different topics, and as we strike out
longer and longer random paths, the topic distribution of pages
collected from these paths start to converge (\S\ref{sec-base-converge}).
This gives us strong circumstantial evidence that the Web has a
well-defined background topic distribution, even though we cannot
directly measure the fidelity of our distribution w.r.t.\ the `whole'
Web.


\paragraph{Degree distribution restricted to topics:}
Once we have some confidence in collecting samples which are faithful
to the Web's topic distribution, we can measure the degree
distribution of pages belonging to each topic-based community.  We
observe (\S\ref{sec-base-degree}) 
that the degree distribution for many topics follow the power
law, just like the global distribution reported by Br\"oder and others.
We offer a heuristic explanation for this observation.

\paragraph{How topic-biased are breadth-first crawls?}
Several production crawlers follow an approximate breadth-first
strategy.  A breadth-first crawler was used to build the Connectivity
Server \cite{BharatBHKV1998connect,BroderKMRRSTW2000bowtie} at
Alta~Vista.  Najork and Weiner \cite{NajorkW2001crawl} report that a
breadth-first crawl visits pages with high PageRank early, a valuable
property for a search engine.  A crawl of over 80 million pages at the
NEC Research Institute
broadly follows a breadth-first policy.  However, can we be sure that
a breadth-first crawl is faithful to the Web's topic distribution?  We
observe that any bias in the seed URLs persists up to a few links
away, but the fidelity does get slowly better (\S\ref{sec-base-bfs}).


\paragraph{Representation of topics in Web directories:}
Although we use Web directories to derive our system of topics, they
need not have fair representation of topics on the Web.  By studying
the difference between the directory and background 
distributions (\S\ref{sec-base-rep}), we
can spot Web topics and communities that have an unexpectedly low (or
high) representation on the directory, and try to understand why.


\paragraph{Topic convergence on directed walks:}
We also study (\S\ref{sec-wander})
page samples collected from ordinary random walks that
only follow hyperlinks in the forward
direction~\cite{HenzingerHMN2000sample}.  We discover that these
ordinary walks do not lose the starting topic memory as quickly as
undirected walks, and they do not approach the background distribution
either.  Different communities lose the topic memory at different
rates.  These phenomena give us valuable insight into the success of
focused crawlers
\cite{ChakrabartiVD1999focus1,DiligentiCLGG2001context,RennieM1999reinforce}
and the effect of topical clusters on Google's PageRank algorithm
\cite{BrinP1998anatomy,PageBMW1998pagerank}.


\paragraph{Link-based vs.\ content-based Web communities:}
We extend the above measurements to construct a \emph{topic citation}
matrix in which entry $(i,j)$ represents the probability that a page
about topic $i$ cites a page about topic 
$j$ (\S\ref{sec-cite}).  The topic citation
matrix has many uses that we will discuss later.  Typically, if the
topics are chosen well and if there is topical clustering on the Web,
we expect to see heavy diagonals in the matrix and small off-diagonal
probabilities.  Prominent off-diagonal entries signify human judgment
that two topics adjudged different by the taxonomy builders are found
to have connections endorsed by Web content writers.  Such
observations may guide the taxonomy builders to improve the structure
of their taxonomies.


\subsection{Limitations}

A fixed taxonomy with coarse-grained topics has no hope of capturing
all possible information needs of Web users, and new topics emerge on
the Web all the time.  Another concern is the potentially low accuracy
of a state-of-the-art text classifier on even such broad topics.
Despite these limitations, we believe that data collected using a
fixed classification system and an imperfect classifier can still be
valuable.

We argue that new topics and communities, while numerous, are almost
always extensions and specializations of existing broad classes.
Topics at and near the top levels of Yahoo! and Dmoz change very
rarely, if at all.

As long as the automatic classifier gives us a significant boost in
accuracy beyond random choice, we need not be overly concerned about
the absolute accuracy (although larger accuracy is obviously better).
By using a held-out data set, we can estimate which topics are
frequently confused by our classifier, and suitably discount such
errors when interpreting collected data.

Recent hypertext classification algorithms
\cite{ChakrabartiDI1998hyperclass} may reduce the error by using
hyperlink features, but this may produce misleading numbers: first the
classifier uses link structure to estimate classes, then we correlate
the class with link structure---the results are likely to show
artificial coupling between text and link features.

A classifier also acts as a dimensionality reduction device and makes
the collected data more dense.  Since we use a few hundred classes but
Web documents collectively use millions of term features, estimating
class distributions is easier than estimating term distributions.



\section{Building blocks\label{sec-algo}}

\subsection{Sampling Web pages\label{sec-intro-pagerank}}

Sampling approximately uniformly at random (uar) is a key enabling
technology that we use in this work, and we review sampling techniques
in some detail here.  The Web graph is practically infinite, because URLs
can embed CGI queries and servers can map static-looking URLs to
dynamic content.  We can discuss sample quality only in the context of
a finite, static graph.  
A small set of strict rules often suffices to make the
Web graph effectively finite, even if unspeakably large.  E.g.,
\begin{itemize}
\item URLs with substrings in the following list are disallowed:
\verb|?|, \verb|cgi-bin|, \verb|&|
\item URLs with more than some maximum number of path components
(counted by slashes) are disallowed
\item URLs are permitted to have some maximum number of characters.
\end{itemize}
Suppose the Web is static and we start from a given set of URLs and
crawl as far as we can, ignoring URLs which do not satisfy the above
rules, collecting a graph $G$ in the process.  Having fetched all of
$G$, we can easily sample a URL uar from $G$.  The key question is,
can we sample a URL uar from $G$ while fetching a subgraph much
smaller than $G$?

\paragraph{PageRank-based random walk:}
Henzinger et al.~\cite{HenzingerHMN2000sample} proposed an early
approach to this problem using PageRank
\cite{BrinP1998anatomy,PageBMW1998pagerank}.  The PageRank prestige
$\pi(u)$ of a node $u$ in $G$ is defined as the relative rate of
returning to node $u$ during an infinite random walk on $G$, where
each step from a node $v$ is taken as follows: with probability $d$
($0<d<1$) we jump to a random node in $G$, and with probability $1-d$
we follow an outlink from $v$ uar.  The steady state distribution
$\mathbf{\pi}$ of this walk is called the PageRank.  Henzinger et al.\
first performed a PageRank-style walk for some steps, and then
\emph{corrected} the bias by sampling the visited nodes with
probability inversely proportion to their $\pi$ scores.
Rusmevichientong et al.~\cite{RusmevichientongPLL2001sample} 
have enhanced this algorithm
and proved that uniform sampling is achieved asymptotically.

\paragraph{The Bar-Yossef random walk:}
An alternative to the biased walk followed by the correction is to
modify the graph so that the walk itself becomes unbiased.  Bar-Yossef
et al.~\cite{BarYossefBCFW2000sample} achieve this by turning the Web
graph into an \emph{un}directed, \emph{regular}\footnote{All
nodes have equal degree in a regular graph.} graph, for which the
PageRank vector is known to have identical values for all nodes.  The
links are made undirected by using the \verb|link:...| backlink query
facility given by search engines.  This strategy parasites on other
people having crawled large sections of the Web to find the
backlinks, and is not guaranteed to find all backlinks.
This will introduce some initial \emph{bias} in the sample towards
pages close to the starting point of the walk.  Unfortunately,
there is no easy way around this bias until and unless
hyperlinks become bidirectional entities on the 
Web~\cite{ChakrabartiGM1999back}.  However we can assess the quality
of the samples through other means.
The graph is made regular by adding sufficient numbers of
self-loops at each node; see~\S\ref{sec-base}.

We use a variant of Bar-Yossef's walk, with random jumps thrown in with
a small probability, which in our empirical experience gave us faster
convergence.  We call this the \textsc{Sampling} walk, whereas the
PageRank walk with $d=0$ is called the \textsc{Wander} walk.



\subsection{Taxonomy design and document
classification\label{sec-intro-classify}}

We downloaded from the Open Directory (\url{http://dmoz.org}) an RDF
file with over \textbf{271,954} topics arranged in a tree hierarchy with
depth at least \textbf{6}, containing a total of about \textbf{1,697,266}
sample URLs.
Since the set of topics was very large and many topics had scarce
training data, we pruned the Dmoz tree to a manageable frontier 
as described in \S3.1 of our companion paper in these 
proceedings~\cite{ChakrabartiPS2002focus}.
%
%
The pruned taxonomy had \textbf{482} leaf nodes and a total of
\textbf{144,859} sample URLs.  Out of these we could successfully fetch
about \textbf{120,000} URLs.

For the classifier we used the public domain BOW toolkit and the
Rainbow naive Bayes (NB) classifier created by McCallum and others
\cite{Bow1998}.  Bow and Rainbow are very fast C implementations which
let us classify pages in real time as they were being crawled.
Rainbow's naive Bayes
learner is given a set of training documents, each labeled with one of
a finite set of classes/topics.  A document $d$ is modeled as a
multiset or bag of words, $\{\langle t, n(d,t)\rangle\}$ where $t$ is
a term/word/token/feature.  The prior probability $\Pr(c)$ is the
fraction of training documents labeled with class $c$.  The NB model
is parameterized by a set of numbers $\theta_{c,t}$ which is roughly
the rate of occurrence of term $t$ in class $c$, more exactly,
$(1+\sum_{d\in D_c}n(d,t))/(|T|+\sum_{d,\tau}n(d,\tau))$, where $D_c$
is the set of documents labeled with $c$ and $T$ is the entire
vocabulary.  The NB learner assumes independence between features, and
estimates
\begin{eqnarray}
\Pr(c|d) &\propto& \Pr(c)\Pr(d|c) \quad\approx\quad
\Pr(c) \prod_{t\in d} \theta_{c,t}^{n(d,t)}.
\end{eqnarray}
Nigam et al.\ provide further details \cite{McCallumN98}.

\section{Background topic distribution\label{sec-base}}

In this section we seek to characterize and estimate the distribution
of topics on the Web, i.e., the fractions of Web pages relevant to a
set of given topics which we assume to cover all Web content.

We will use only \textsc{Sampling} walks here;
\textsc{Wander} walks will be used in \S\ref{sec-wander}.
Bar-Yossef et al.'s random walk has a few limitations.  Apart from the
usual bias towards high indegree node and nodes close to the starting
point, it may find convergence elusive in practice if snared in
densely linked clusters with a sparse egress.  One example of such a
graph is the ``lollipop graph'' which is a completely-connected
$k$-clique with a `stick' of length $k$ dangling from one node
belonging to the clique.  It is easy to see that a walk starting on
the stick is doomed to enter the clique with high probability, whereas
getting out from the clique to the stick will take a long, long
time~\cite{MotwaniR1995random}.


Unfortunately, lollipops and near-lollipops are not hard to find on
the Web: \url{http://www.amazon.com/}, \url{http://www.stadions.dk/} and
\url{http://www.chipcenter.com/} are some prominent examples.  Hence we
add a PageRank-style random jump parameter to the original Bar-Yossef
algorithm \cite{BarYossefBCFW2000sample}, set to 0.01--0.05 throughout
our experiments, i.e., with this probability at every step, we jump
uar to a node visited earlier in the \textsc{Sampling} walk.  Berg
confirms \cite{Berg2001walk} that this improves the stability and
convergence of the \textsc{Sampling} walks.

\subsection{Convergence\label{sec-base-converge}}

Bar-Yossef et al.\ showed that the samples collected by a
\textsc{Sampling} walk have degree distributions that converge quickly
(within a few hundred distinct page fetches) to the ``global'' degree
distribution (obtained from the Internet Archive).  It would be
interesting to see if and how convergence properties generalize to
other page attributes, such as topics.

\begin{figure}
\begin{center}
\iftth
  \tthfig{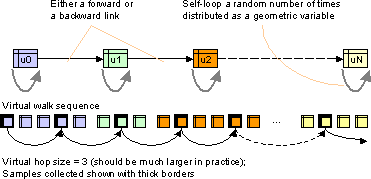}
\else
  \tthfig{\includegraphics*[width=\hsize]{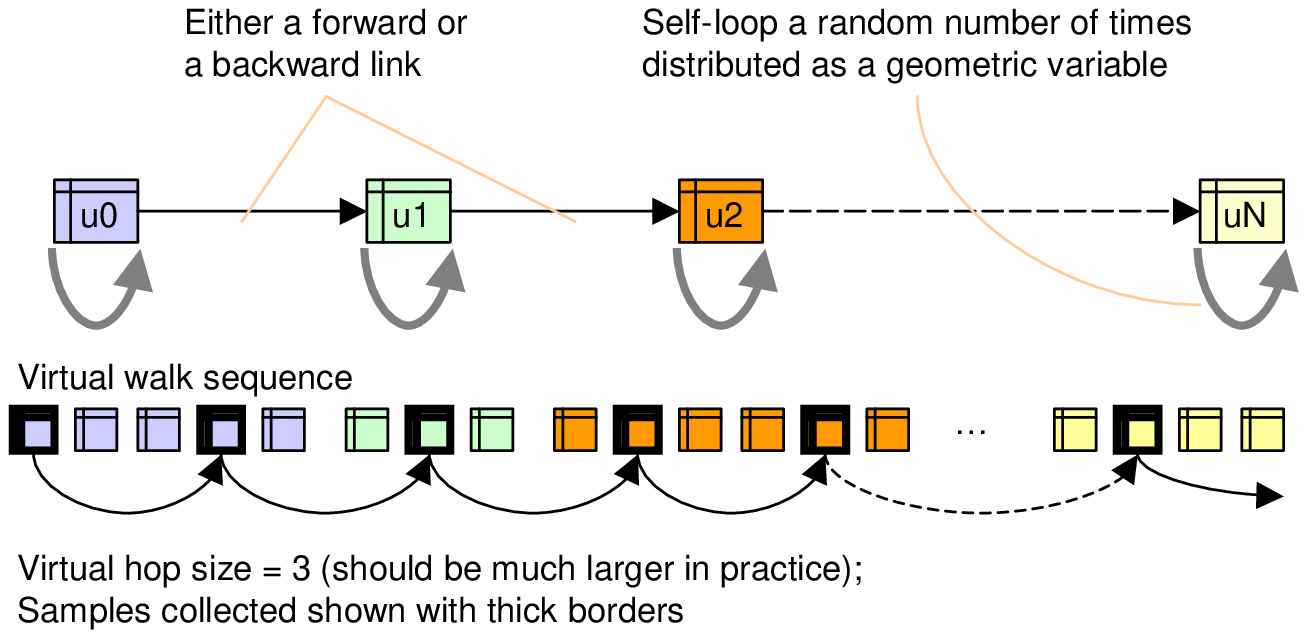}}
\fi
\caption{Using a virtual stride on the Bar-Yossef walk to 
derive a near-random sample.}  \label{fig-virtual}
\end{center}
\end{figure}

Our basic experimental setup starts with two URLs $u_0$ and $v_0$,
generally about very different topics.  We now execute the following
steps:
\begin{enumerate}
\item
Perform separate \textsc{Sampling} walks starting from each of them.  In
our implementation we do not model the self-loops for efficiency, so
we get a what we call a ``physical walk'' where successive URLs are
generally different.
\item
We estimate an upper bound to the maximum degree $M$ of 
any node in the Web.  We can pick a very loose upper bound, such
as 10,000,000---this will increase the number of self-loops, but have 
no effect on the efficiency or the quality of the sample.
The self-loop probability of a
node $u$ with total (in+out) degree $d_u$ is set to $1-d_u/M$.
\item
We turn the physical walk into a ``virtual walk'' by self-looping at
each URL $u$ for a random number of times, distributed as a geometric
random variable with mean $d_u/M$.  (See
\figurename~\ref{fig-virtual}.)  A geometric random variable $X$ with
mean $\theta$ has value $x$ with probability
$\Pr(X=x) = \theta(1-\theta)^{x-1}$, for $x=1,2,\ldots$.

\item
Ideally, we should use one \textsc{Sampling} walk for collecting each
sample page (i.e., keep only the last page
reached in every \textsc{Sampling} walk), 
but walking is expensive (mainly because of backlink queries
which need to be polite to the search service).  Therefore we pick a
sufficiently large (virtual) stride $k$ over which we hope ``memory is
lost'' on the virtual walk, and collect URL samples every $k$ hops
from the virtual walk.  We play with a range of $k$s to guard against
too small a choice.
\end{enumerate}


In our first experiment, we pick some 20 topics from our 482-topic
Dmoz collection and one representative URL from each topic as a
starting point for a \textsc{Sampling} walk.  Each page visited in a
walk is classified using Rainbow and its class histogram as well as
in- and out-neighbors stored in a relational database.  Then we
consider pairs of walks, turn them into virtual walks and sample at a
given stride on the fly.

Suppose we have thus collected two sets of documents $D_1$ and $D_2$.
Each document $d$ has a class probability vector $\mathbf{p}(d) =
(\Pr(c|d)\forall c)$, where $\sum_c\Pr(c|d)=1$.  E.g., if we have only
two topics \verb|/Arts| and \verb|/Sports|, a document may be 
represented by the vector $(0.9,0.1)$.  The topic
distribution of $D_1$ (likewise, $D_2$) is simply the average.
\begin{eqnarray}
\mathbf{p}(D_1) &=& \frac{1}{|D_1|}\sum_{d\in D_1}\mathbf{p}(d).
\end{eqnarray}
This is a form of \emph{soft counting}.  The `hard' analog would be to
assign each $d$ to its highest scoring class and count up the number
of documents assigned to each class.  Lewis \cite{Lewis2000attics}
notes that soft counting gives better estimates than hard counting for
small sample sizes.

We characterize the \textbf{difference} between $\mathbf{p}(D_1)$ and
$\mathbf{p}(D_2)$ as the $L_1$ difference between the two vectors,
$\sum_c|p_c(D_1)-p_c(D_2)|$.  This difference ranges between 0 and~2
for any two probability vectors over the classes.  
We had to avoid the
use of the more well-known Kullback-Leibler (KL) divergence
\begin{eqnarray}
KL(\mathbf{p}(D_1)||\mathbf{p}(D_2)) &=&
\sum_c p_c(D_1)\log\frac{p_c(D_1)}{p_c(D_2)}
\end{eqnarray}
because of two problems: it is asymmetric, and more seriously, it
cannot deal with zero probabilities gracefully.  
The symmetric Jensen-Shannon (JS) divergence \cite{CoverT1991}
also has problems with zeroes.

Bar-Yossef et al.\ found that an undirected random walk touching about
300 physically distinct pages was adequate to collect a URL
sufficiently unbiased to yield a good degree distribution estimate.
We use this number as a guideline to set the virtual hop size so that
at least this number of physical pages will be skipped from one sample
to the next.

\begin{figure}[t]
\begin{center}
\begin{tabular}{c}
\iftth
\tthfig{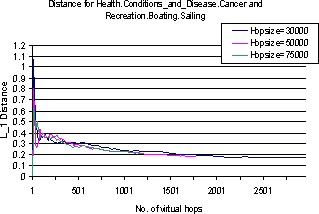} \\
\tthfig{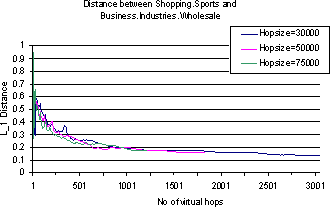} \\
\tthfig{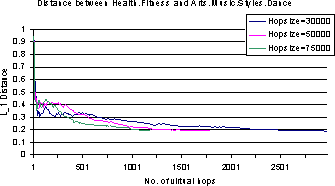}
\else
\tthfig{\includegraphics*[width=\hsize]{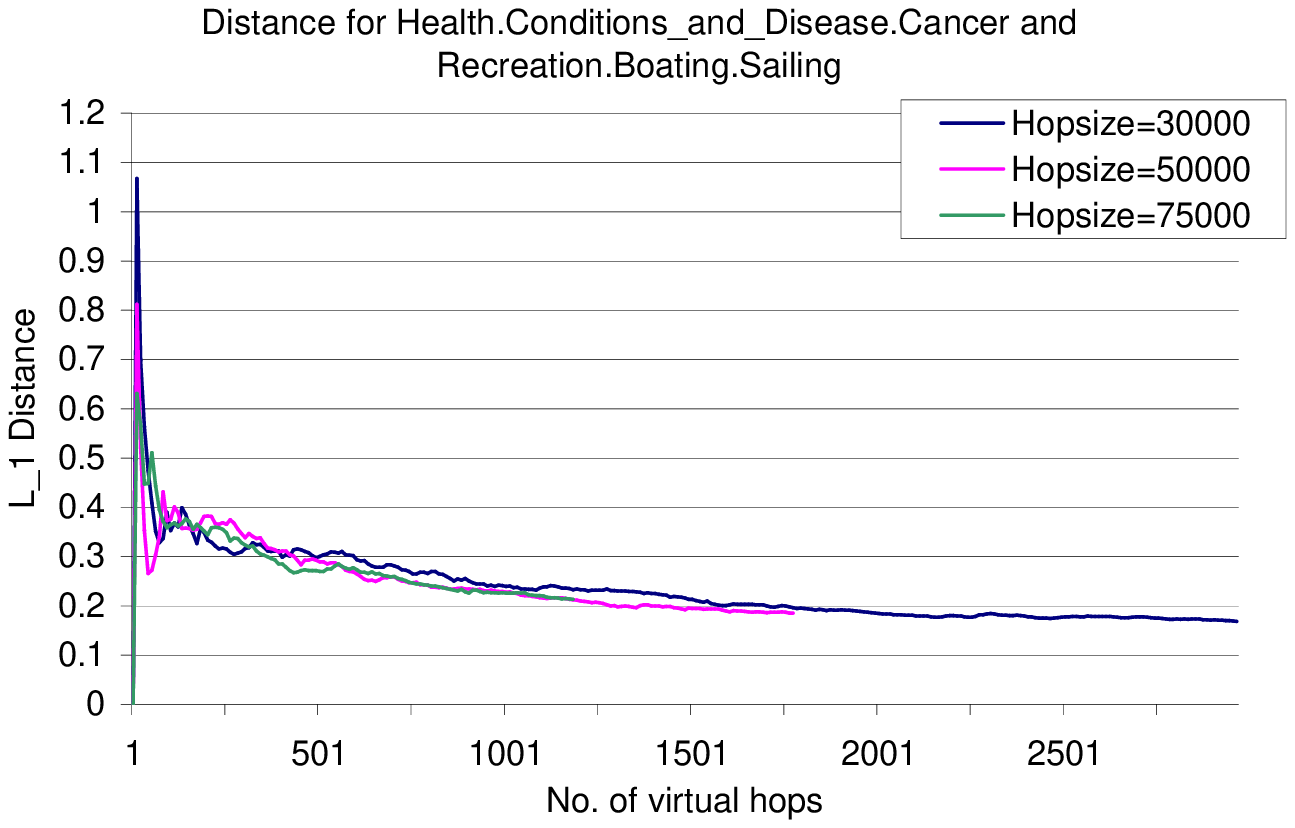}} \\
\tthfig{\includegraphics*[width=\hsize]{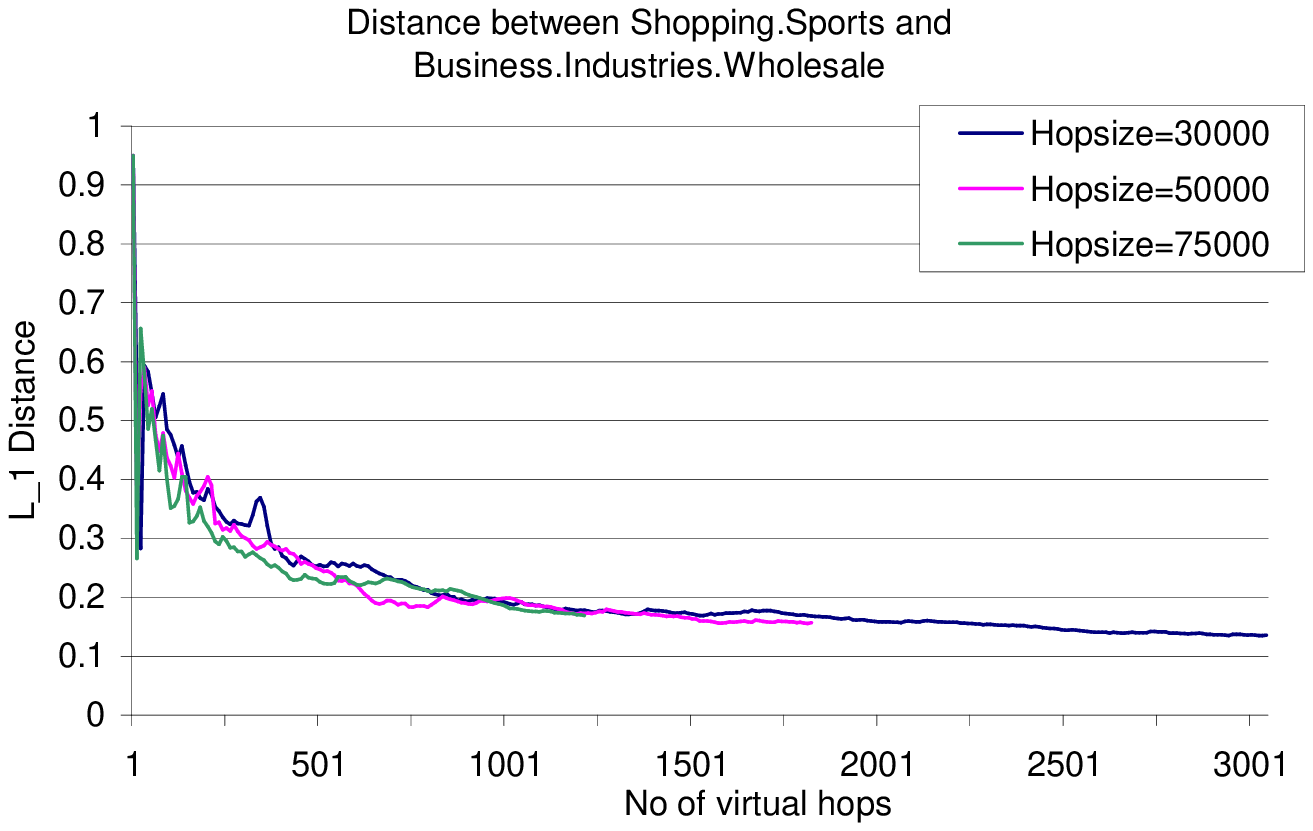}} \\
\tthfig{\includegraphics*[width=\hsize]{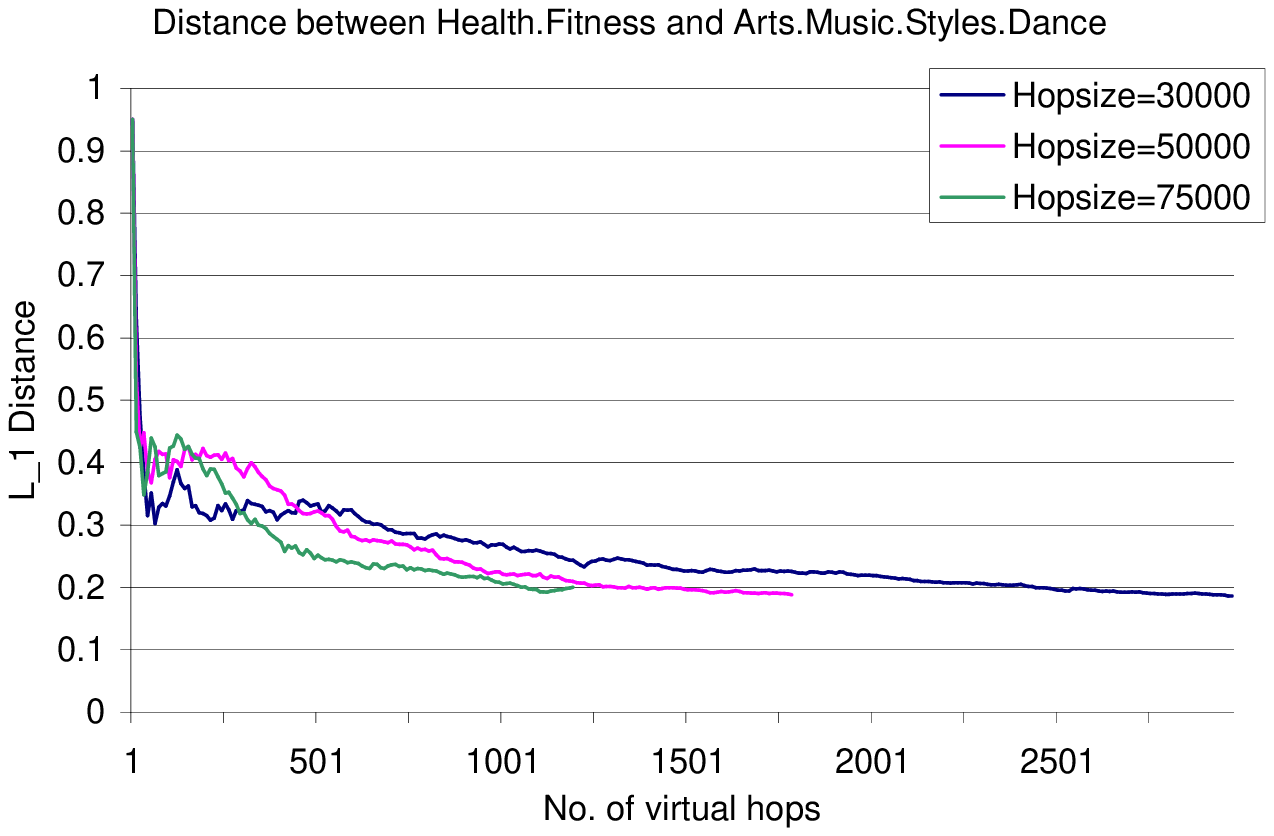}}
\fi
\end{tabular}
\caption{Topic convergence of \textsc{Sampling} walks using the
detailed Dmoz topics.  Starting from several pairs of very distinct
topics, \textsc{Sampling} walks converge to each other (which hints
that it is also the Web's background topic distribution) within a few
thousand virtual hops.  The virtual hop width is varied between 30000
and 75000, the number of virtual hops is the x-axis, and the y-axis is
the $L_1$ distance between the two topic probability vectors.}
\label{fig-sample-diff-detailed}
\end{center}
\end{figure}

\figurename~\ref{fig-sample-diff-detailed} shows pairwise
class distribution differences starting from a few pairs of very
dissimilar topics; the x-axis plots the number of samples drawn from
the respective virtual walks, and different curves are plotted for
different virtual hop sizes.  We never seem to get complete
convergence (distance zero) but the distance does rapidly reduce from
a high of over 1 to a low of about 0.19 within 1000--1500 virtual
hops, which typically includes about 300--400 distinct physical pages.
A larger virtual stride size leads to a slightly faster rate of
convergence, but curiously, all pairs converge within the same
ballpark number of virtual strides.

Convergence of this nature is a stronger property than just a decay in
the similarity between $u_0$ and $u_i$ on a walk $(u_0,u_1,\ldots)$ as
$i$ increases.  It indicates that there \emph{is} a well-defined
background topic distribution and we are being able to approach it
with suitably long \textsc{Sampling} walks.

\begin{figure}[t]
\begin{center}
\begin{tabular}{c}
\iftth
\tthfig{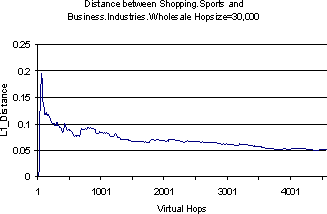} \\
\tthfig{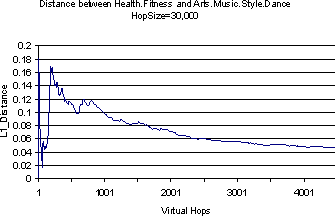}
\else
\tthfig{\includegraphics*[width=\hsize]{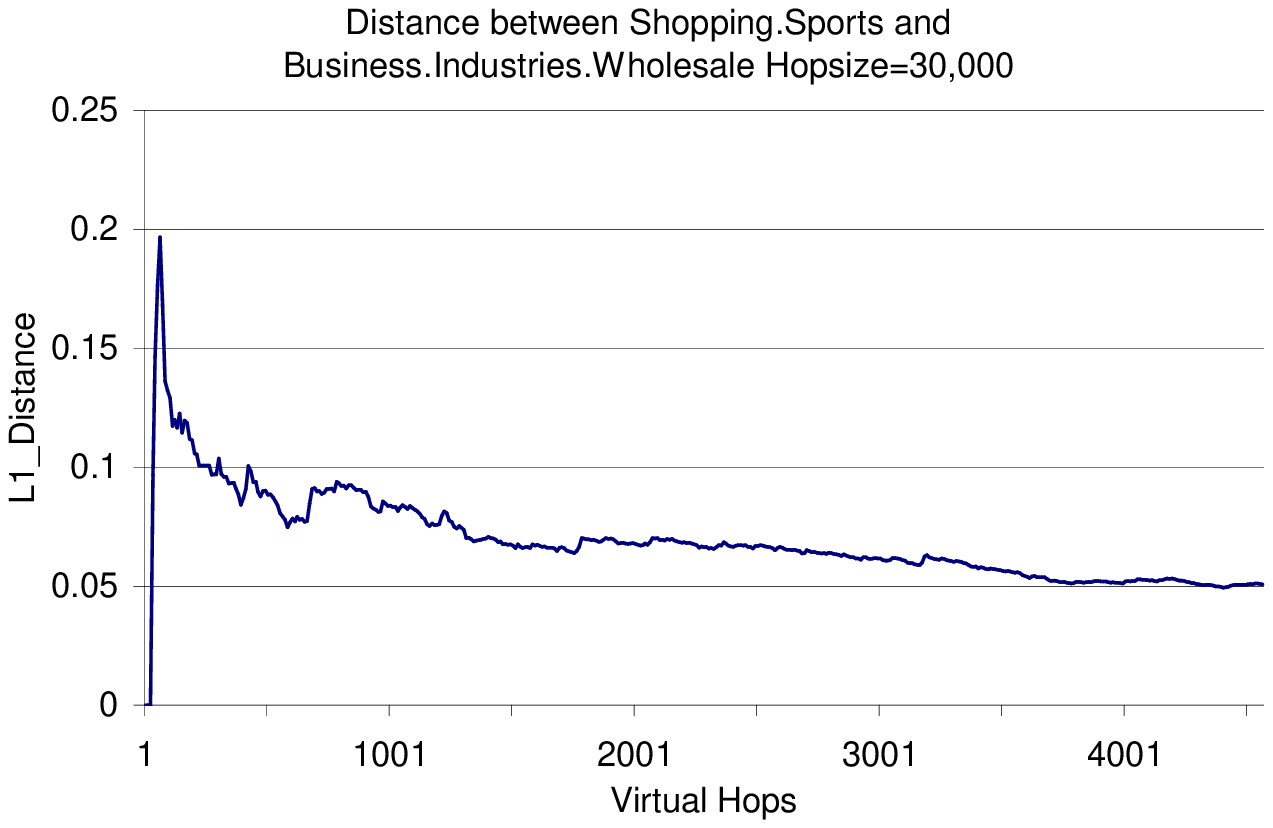}} \\
\tthfig{\includegraphics*[width=\hsize]{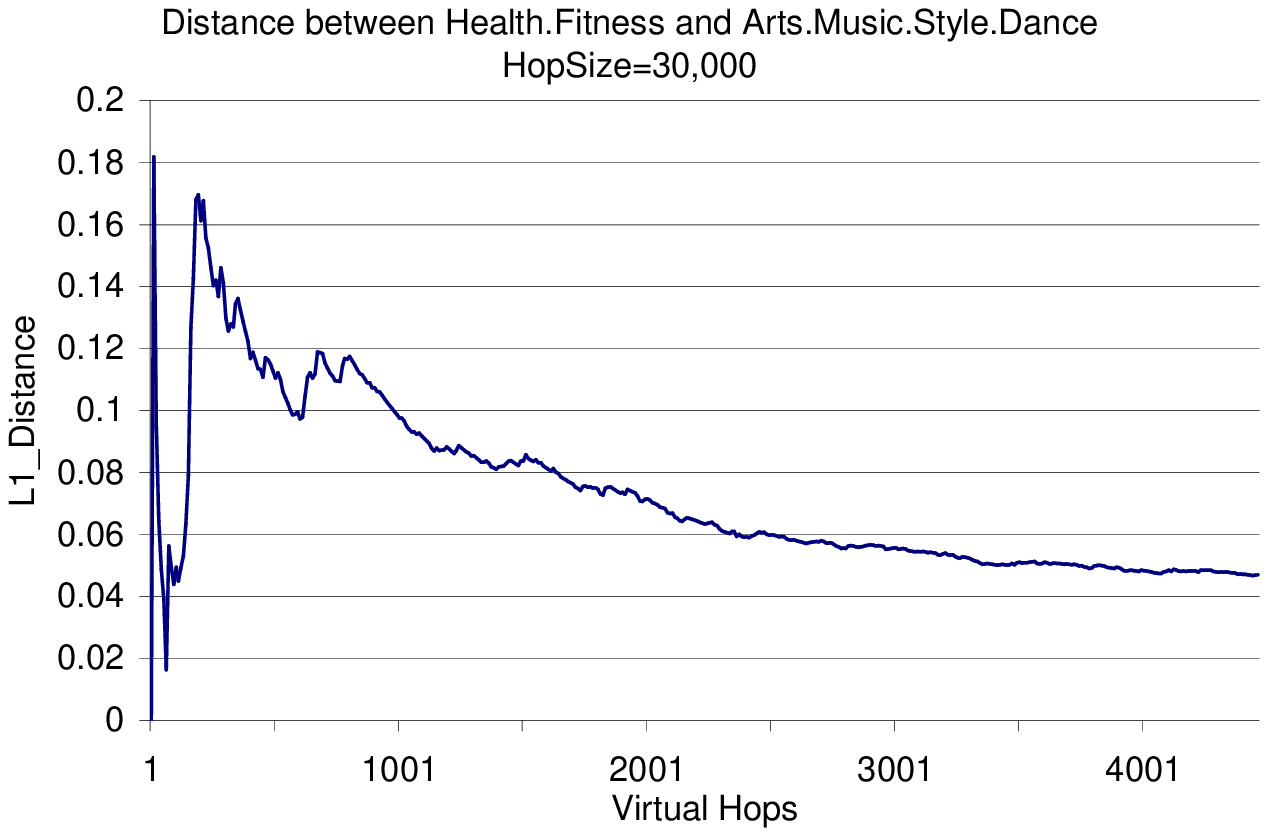}}
\fi
\end{tabular}
\caption{Topic convergence of \textsc{Sampling} walks using
a version of Dmoz topic obtained by coarsening each leaf topic
to its immediate parent.  The number of virtual hops
required is the same order of magnitude as with the detailed topics,
but the $L_1$ inter-walk distance achieved is much lower.}
\label{fig-sample-diff-coarse}
\end{center}
\end{figure}

Obviously, the rate at which the topic probability vectors converge
depends on the granularity of the topic specification, if only because
it will take a larger number of documents to fill up a larger number
of topic buckets adequately to make a reliable reading.  We repeat the
above experiment with a coarser version of Dmoz which is obtained by
lifting all the 482 leaf classes of our Dmoz topic set to their
immediate parents.  \figurename~\ref{fig-sample-diff-coarse} shows the
results.  Because topic bins are populated more easily now, the
distance is already small to start with, about 0.2, and this decreases
to 0.05.  But the number of virtual hops required is surprisingly
resilient, again within the 1000--1500 range.

There are some anomalous drops in distance at a small number of
virtual hops in the graphs mentioned above.  This is because we did
not preserve the page contents during the walks but re-fetched them
later for pages classification.  Some page fetches at small hop count
timed out, leading to the instability.  For larger hop counts the
fraction of timeouts was very small and the result became more stable.

\begin{figure}[ht]
\begin{center}
\iftth
\tthfig{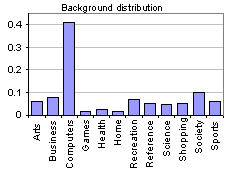}
\else
\tthfig{\includegraphics*{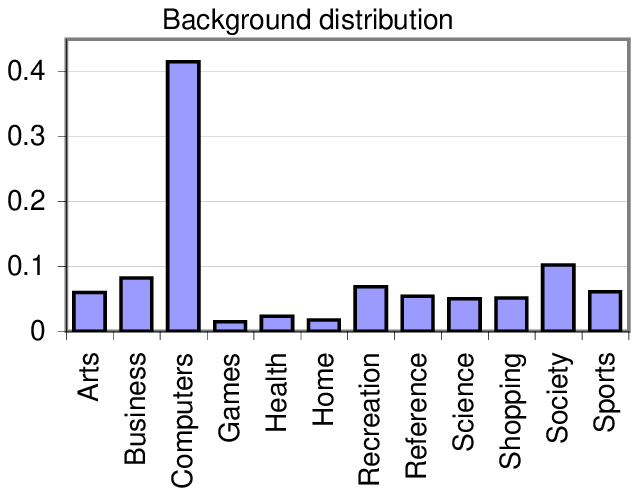}}
\fi
\caption{An estimate of the background distribution across
the 12 top-level topics in our taxonomy.}
\label{fig-back12u}
\end{center}
\end{figure}

In \figurename~\ref{fig-back12u}, we show an estimate of the background
distribution of Web pages into the 12 top-level topics in our
taxonomy.  Computer-related topics lead the show.  This is completely
understandable (if only because of rampant mirroring of software
manuals).

\subsection{The background distribution vs 
breadth-first crawls\label{sec-base-bfs}}

Many production crawlers follow an approximate breadth-first link
exploration strategy which gives them some basic robustness against
overloading a small set of servers or going depth-first into a
bottomless ``spider pit''.  The crawler which populates Alta~Vista's
Connectivity Servers follows largely a breadth-first
strategy~\cite{BharatBHKV1998connect,BroderKMRRSTW2000bowtie}.  Najork
and Weiner \cite{NajorkW2001crawl} have demonstrated that a
breadth-first crawler also tends to visit nodes with large PageRank
early (because good authorities tend to be connected from many places
by short paths).  A crawler of substantial scale deployed in NEC
Research uses breadth-first scheduling as well.


\begin{figure}[ht]
\begin{center}
\begin{tabular}{|l|r|r|r|}  \hline
Level & Sampled & Unique & Fetched \\ \hline
0	& 50000		& 49274		& 23276		\\
1	& 500000	& 491893	& 45839		\\
2	& 5000000	& 4857239	& 109544	\\
\hline
\end{tabular}
\caption{Some statistics of our sample of 
the three-level breadth-first
NEC crawl.} \label{fig-nec-crawl-stats}
\end{center}
\end{figure}




The NEC crawl was started from URLs taken from Dmoz.  These URLs were
placed in level zero.  We collected URL samples from levels 0, 1 and~2.
Details are shown in \figurename~\ref{fig-nec-crawl-stats}.
\figurename~\ref{fig-nec-results}(a) shows the pairwise distance between
the three levels of the NEC crawl.  The distances are fairly small,
which indicates that the aggregate class distribution drifts quite
slowly as one strikes out from the seed set.  Therefore any
significant bias in the seed set will persist for quite a few levels,
until the frontier size approaches a sizable fraction of the reachable
Web.

\begin{figure}[ht]
\begin{center}
\begin{tabular}{cl}
(a) &
\begin{tabular}{|c|c|r|} \hline 
First URL set & Second URL set & $L_1$ distance \\ \hline
NEC0 & NEC1 & 0.28202 \\
NEC1 & NEC2 & 0.23273 \\
NEC0 & NEC2 & 0.37231 \\
\hline
\end{tabular}
\\
(b)& 
\begin{tabular}{|c|r|}  \hline
URL set & $L_1$ distance from background \\ \hline
NEC0 & 0.65902 \\
NEC1 & 0.55181 \\
NEC2 & 0.58630 \\
\hline
\end{tabular}
\end{tabular}
\caption{The topic composition of a breadth-first crawl changes slowly,
showing small distance between adjacent layers.  Consequently, if the
first layer is biased, the bias persists for some depth.
(a)~Pairwise distances between the first three layers of the NEC crawl.
(b)~Distance between NEC crawl layers and our background estimate.}
\label{fig-nec-results}
\end{center}
\end{figure}

\figurename~\ref{fig-nec-results}(b) shows the distance of the 
NEC collections at the three levels and our approximation to the
background topic distribution.  We see some significant distance
between NEC collections and the background distribution, again
suggesting that the NEC topic distributions carry some bias from the
seed set.  The bias drops visibly from level~0 to level~1 and then
rises very slightly in level~2.  It would be of interest to conduct
larger experiments with more levels.

\begin{figure}
\begin{center} 
\begingroup \tabcolsep1pt \footnotesize
\begin{tabular}{|l|} \hline
\textcolor{blue}{\bfseries Topics OVER-represented in Dmoz}\\
\textcolor{blue}{\bfseries compared to the background}\\ 
\hline
\verb|Games.Video_Games.Genres| \\
\verb|Society.People| \\
\verb|Arts.Celebrities| \\
\verb|Reference.Education.Colleges_and_Universities.North_America...| \\
\verb|Recreation.Travel.Reservations.Lodging| \\
\verb|Society.Religion_and_Spirituality.Christianity.Others| \\
\verb|Arts.Music.Others| \\
\verb|Reference.Others| \\ \hline
\textcolor{blue}{\bfseries Topics UNDER-represented in Dmoz}\\
\textcolor{blue}{\bfseries compared to the background} \\ 
\hline
\verb|Computers.Data_Formats.Markup_Languages| \\
\verb|Computers.Internet.WWW.Searching_the_Web.Directories|\\
\verb|Sports.Hockey.Others| \\
\verb|Society.Philosophy.Philosophers| \\
\verb|Shopping.Entertainment.Recordings| \\
\verb|Reference.Education.K_through_12.Others| \\
\verb|Recreation.Outdoors.Camping| \\ \hline
\end{tabular}
\endgroup
\caption{Some of the largest discrepancies between
the Web's background topic distribution and our selection from Dmoz.}
\label{fig-deviation}
\end{center}
\end{figure}

\subsection{Faithful representation of topics in Web 
directories\label{sec-base-rep}}

Many Web users implicitly expect topic directories to be a microcosm
of the Web itself, in that pages of all topics are expected to be
represented in a fair manner.  Reality is more complex, and commercial
interests play an important role in biasing the distribution of
content cited from a topic directory.  Armed with our sampling and
classification system, we can easily make judgments about the biases
in topic directories, and locate topics which are represented out of
proportion, one way or another.


The $L_1$ distance between the Dmoz collection and the background
topic distribution is quite high, 1.43, which seems to indicate that
the Dmoz sample is highly topic biased.  Where are the biases?  We
show some of the largest deviations in
\figurename~\ref{fig-deviation}.  In continuing work we are testing
which among these are statistically significant.
This is not a direct statement about DMoz, because our sample of DMoz
differs from its original topic composition.  Over-representation
could be caused by our sampling, but the under-represented topics
are probably likewise under-represented in DMoz.  In any case, it
is clear that comparisons with the background distribution gives us 
a handle on measuring the representativeness of topic directories.


\subsection{Topic-specific degree distributions\label{sec-base-degree}}

Several researchers have corroborated that the distribution of degrees
of nodes in the Web graph (and many social networks in general
\cite{FaloutsosFF1999powerlaw,WassermanF1994social}) asymptotically follow a
\emph{power law} distribution
\cite{Adamic00-inlinks,BroderKMRRSTW2000bowtie,RaviKumarRRT1999trawling}:
the probability that a randomly picked node has degree $i$ is
proportional to $1/i^x$, for some constant `power' $x>1$.  The powers
$x$ for in- and out-degrees were estimated in 1999 to be about 2.1 and
2.7, respectively, though the fit breaks down at small $i$, especially
for out-degrees \cite{BroderKMRRSTW2000bowtie}. Barab\'{a}si and
Albert \cite{Barabasi99c} gave an early generative model called
``preferential attachment'' and an analysis for why millions of
autonomous hyperlinking decisions distributed all over the Web could
lead to a power-law distribution.  Dill et al.
\cite{DillRMRST2001fractal} showed that subgraphs selected from the
Web as per specific criteria (domain restriction, occurrence of a
given keyword, etc.) also appear to satisfy power-law degree
distributions. Pennock et al.~\cite{PennockLFLG2001winner} found that
certain topic-specific subsets of the web diverge markedly from
power-law behavior at small $i$, though still converge to a power law
for large $i$; the authors explain these observed distributions using
an extension of Barab\'{a}si's model.

\begin{figure}[ht]
\begin{center}
\begin{tabular}{c}
\iftth
\tthfig{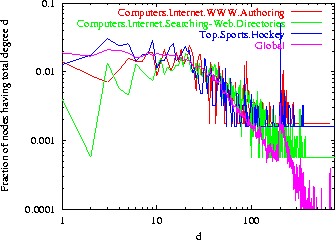} \\
\tthfig{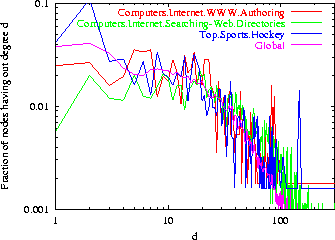}
\else
\tthfig{\includegraphics*[width=\hsize]{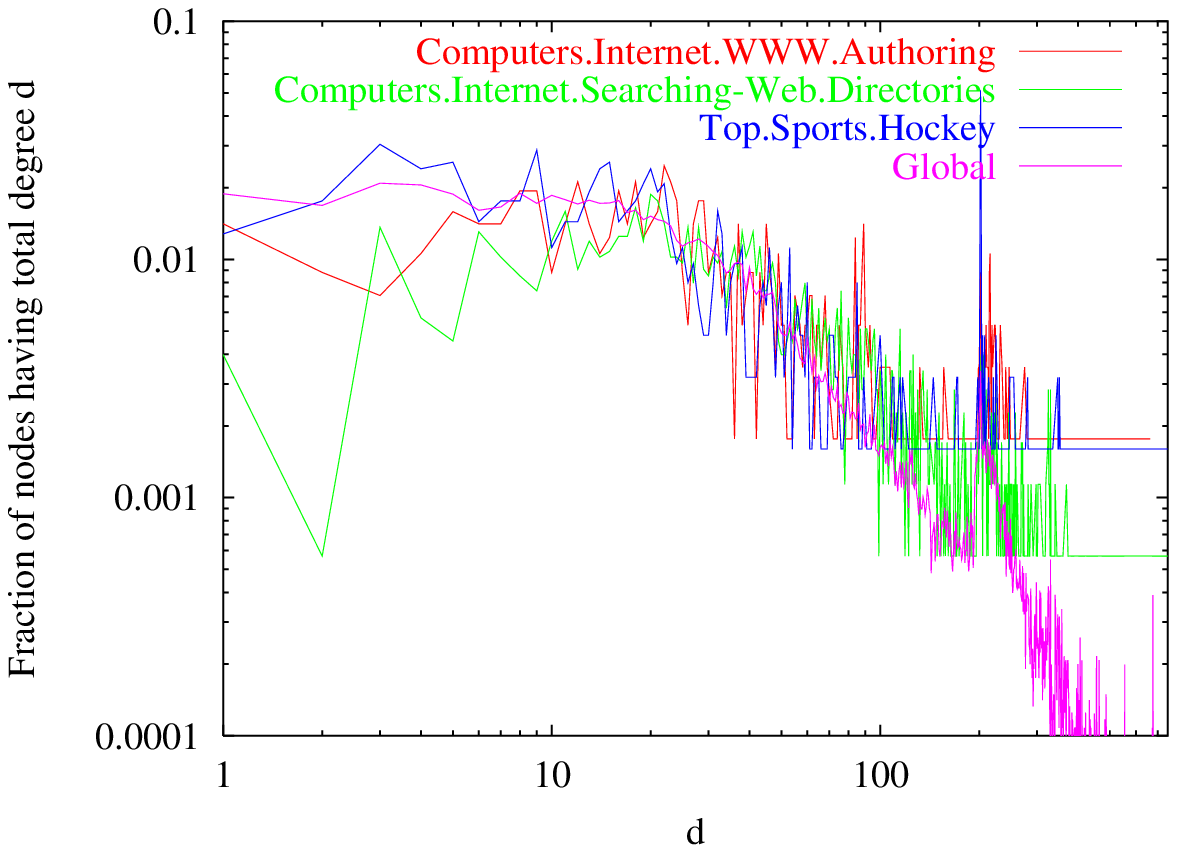}} \\
\tthfig{\includegraphics*[width=\hsize]{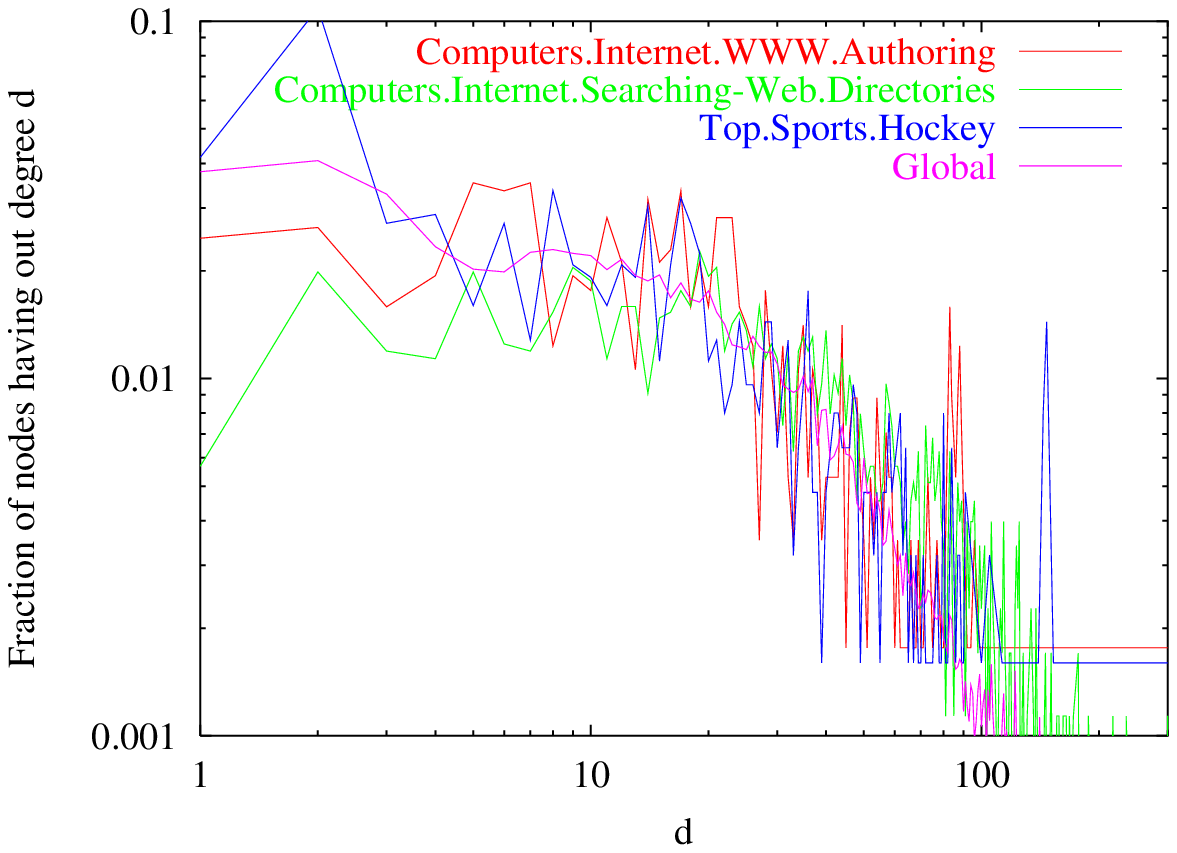}}
\fi
\end{tabular}
\caption{Total degree and out-degree, restricted to a random sample from
various topics, also show a power-law behavior just like degree over
all topics, at least for degree greater than about ten.  The distribution
of well-connected communities like Web directories is shifted slightly
to the right (larger degree) than relatively less connected communities
like Hockey.  Anomalies near 100--200 links are due to 
link spamming.}
\label{fig-degree-topic}
\end{center}
\end{figure}

We conduct more general tests for power law behavior across a broad
array of topics: if we fix a topic and measure the degree of nodes
relevant to that topic, will the resulting degree distribution also
follow a power law?  We can use the same soft-counting technique to
answer this question.  Using a
\textsc{Sampling} walk, we derive a sample $D$ of pages.  If a page
$d$ has degree $\Delta_d$ and class vector $\mathbf{p}(d)$, it
contributes a degree of $\Delta_d\,p_c(d)$ to class $c$.
Note that $\Delta_d$ includes all links incident on~$d$.

\figurename~\ref{fig-degree-topic} shows that topic-specific
degree distributions \emph{also} follow the power law over many orders
of magnitude.  We cannot explain this by claiming that ``social
networking within a topic mimics social networking on the Web at
large'', because $\Delta_d$ includes off-topic links, for which
there is no known analysis of social networking processes.

Log-log plots of degree distributions for various topic 
look strikingly similar at first glance, but a
closer examination shows that, for example, pages about Web
directories generally have larger degree than pages about hockey,
which matches our knowledge of the Web.
We observe that the degree distribution restricted to members of a
specific topic have a power law tail, but with a significant
divergence from power law at small numbers of links, in agreement with
the Pennock et al.\ findings
\cite{PennockLFLG2001winner}, and 
in contrast to the global in-degree distribution which is nearly a
pure power law~\cite{BroderKMRRSTW2000bowtie}.


An empirical result of Palmer and Steffan \cite{PalmerS2000powerlaw}
may help explain why we would expect to see the power law upheld by
pages on specific topics.  They showed through experiments
that the following simple ``80-20'' random graph generator fits
power-law degree distributions quite well:
\begin{enumerate}
\item Assume for simplicity that the number of nodes $N$ is a power
of~2, and $M$ edges are desired.
\item Partition the graph into two node sets $V_1$ and $V_2$
each of size $N/2$.  Let fraction $p_{ij}$ of edges go from some node
in $V_i$ to some node in $V_j$, $i,j=1,2$.  $P=(p_{ij})$ are provided
as parameters. The idea is to favor intra-community links by having
$p_{11}$ and $p_{22}$ larger than $p_{12}$ or $p_{21}$, hence the name
``80-20''.
\item
Using $P$, pick one of the four possible types of edges.
\item
Recurse within the $1/4$ of edges that are consistent with the above
choice; continue recursing with the same parameters $P$ until a
specific edge is materialized (i.e., until the number of remaining
nodes equals 1 or 2).
\item 
Repeat until $M$ edges are added to the graph.
\end{enumerate}
That this ``fractal'' style of construction produces graphs following
power-law degree distributions hints that some of the recursive
diagonal blocks in the adjacency matrix might well represent
topic-specific subgraphs.  In other words, the construction of Palmer
and Steffan seems to indicate that if we constructed separate,
topic-specific subgraphs according to their recipe, and used another
$P$ matrix to drive the formation of links across communities,
the power-law degree distribution would be preserved.
(The inter-community linkage would just be the outermost level
of their recursion.)  Further simulation and/or analysis is needed
to confirm this theory.

\section{Topical locality and link-based 
prestige ranking\label{sec-wander}}

\begin{figure}[th]
\begin{center}
\iftth
\tthfig{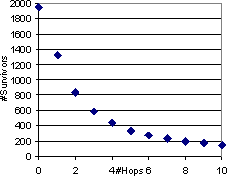}
\else
\tthfig{\includegraphics*[width=.65\hsize]{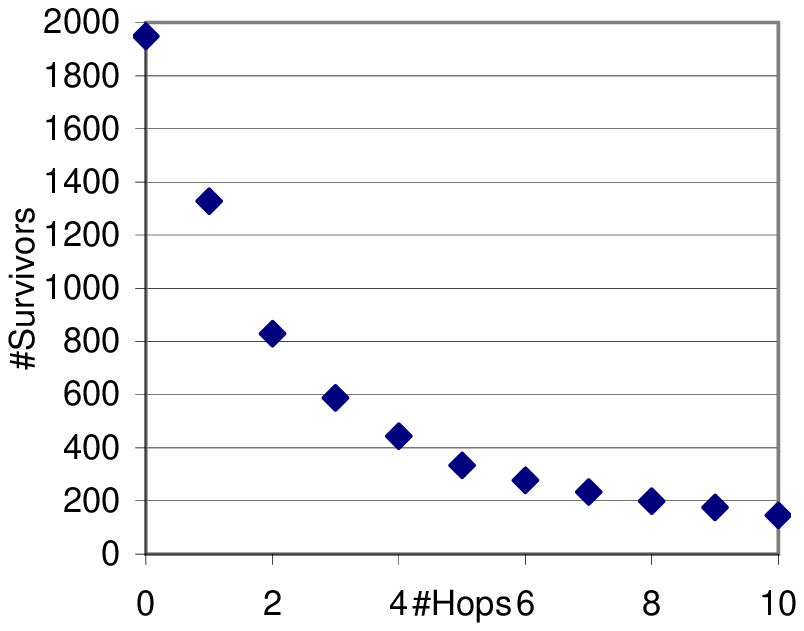}}
\fi
\caption{Very few \textsc{Wander} walks survive for long,
making it difficult to compare topic distributions for diverse
walk radii as with \textsc{Sampling} walks.  This may explain why
the random jump paradigm is so important in PageRank.}
\label{fig-wander-survivor}
\end{center}
\begin{center} \tabcolsep0pt
\begin{tabular}{cc}
\iftth
\tthfig{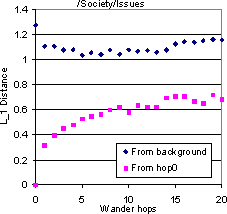} &
\tthfig{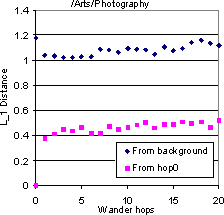} \\
\tthfig{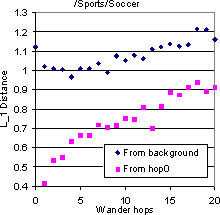} &
\tthfig{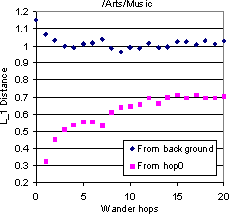}
\else
\tthfig{\includegraphics*[width=.5\hsize]{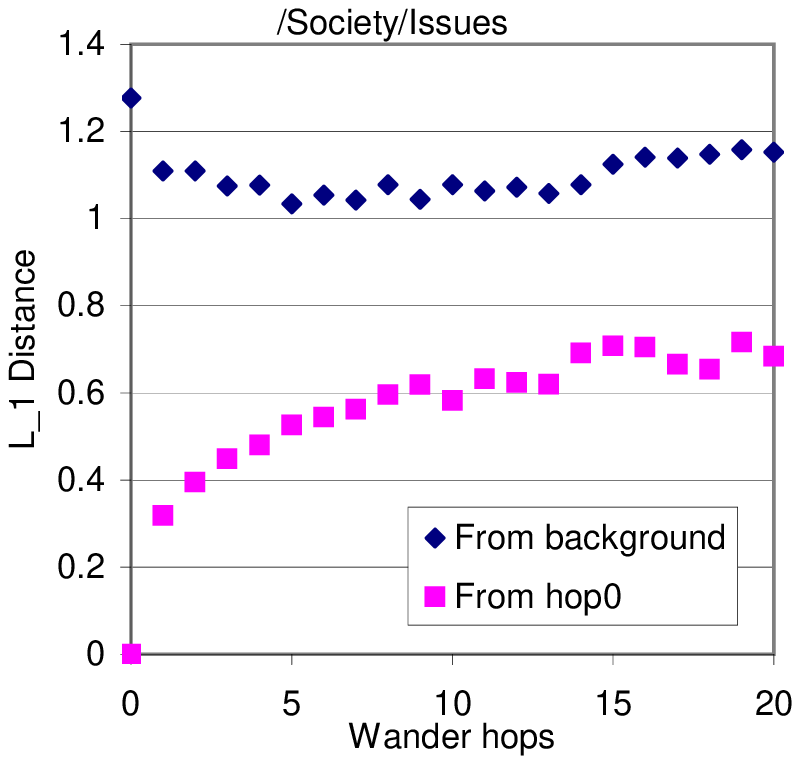}} &
\tthfig{\includegraphics*[width=.5\hsize]{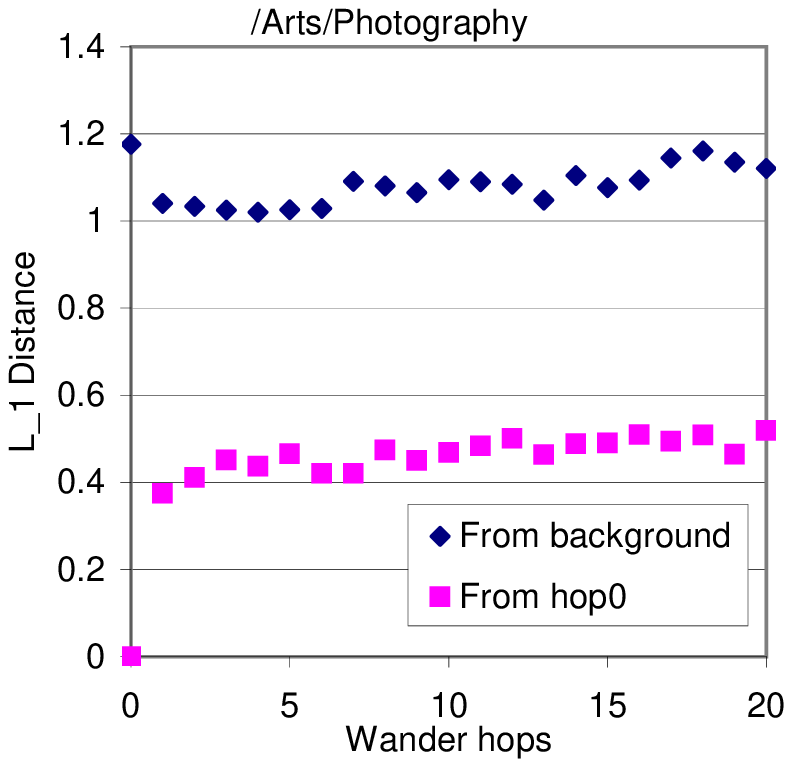}} \\
\tthfig{\includegraphics*[width=.5\hsize]{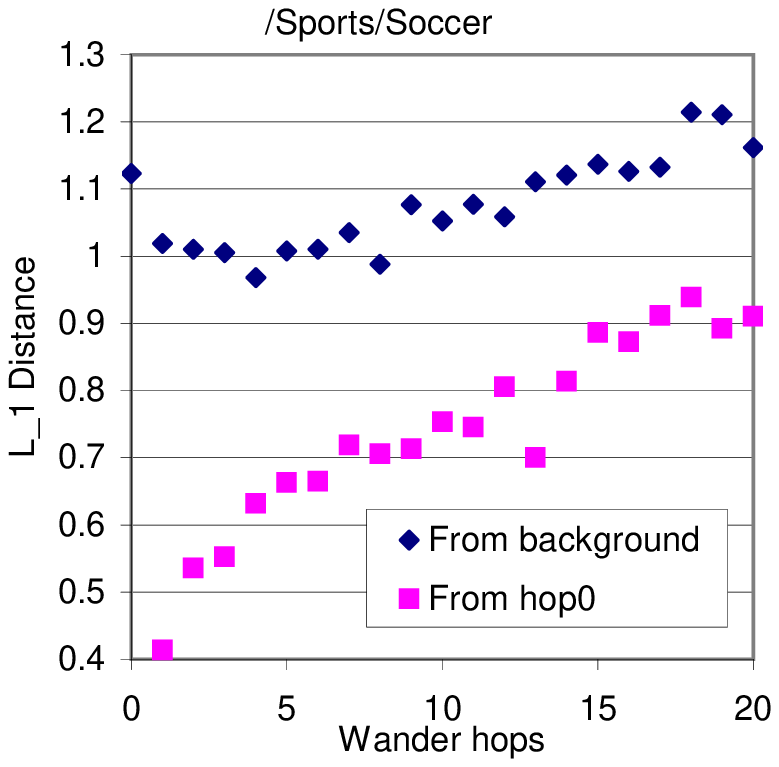}} &
\tthfig{\includegraphics*[width=.5\hsize]{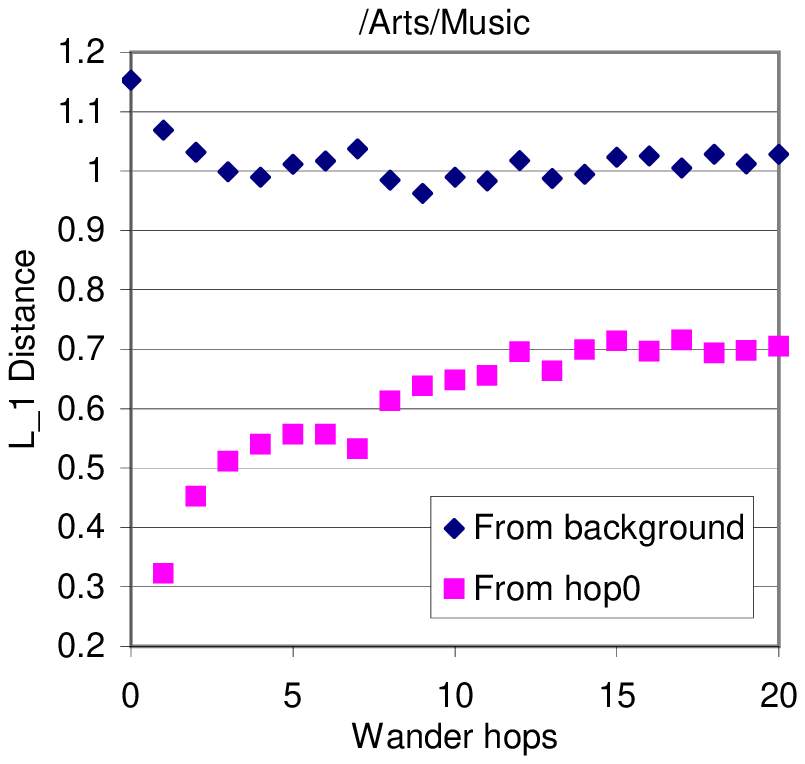}}
\fi
\end{tabular}
\caption{Forward walks without jumps retain topical locality
well beyond the inverse of the jump parameter $d$ in PageRank.  The
distance from the reference background topic distribution changes
slowly (red squares).  Also, in stark contrast to \figurename{}s
\ref{fig-sample-diff-detailed} and \ref{fig-sample-diff-coarse},
there is no sign of convergence to the background distribution
(blue diamonds).}
\label{fig-wander-background}
\end{center}
\end{figure}

In this section we use the \textsc{Wander} walk to see how fast the
memory of the topic of a starting page fades as we take random
\emph{forward} steps along HREFs.  (No backlinks or self-loops 
are permitted.)  It is actually quite difficult in practice to sustain
forward walks.  \figurename~\ref{fig-wander-survivor} shows that if we
start a large set of \textsc{Wander} walks, very few survive with each
additional hop, owing to no outlink, broken outlinks, or server
timeout.  Note that this experiment is different from the study
of the NEC crawler, because here, only \emph{one} random outlink is
explored from each page, whereas the NEC crawler tried to fetch as
many outlinks in every subsequent level as possible.

\subsection{Experiments and results}

At this point we have a reference background topic distribution.  We
consider several classes in DMoz.  Because forward walks die fast, we
picked some Dmoz classes which were very well-populated, say with more
than 10000 URLs.  We start one \textsc{Wander} walk from each URL.  The
starting pages are at distance~0.  For each topic, we collect all the
pages $D_i$ found at distance $i$, $i\ge0$, and find the soft class
histogram $\mathbf{p}(D_i)$ as before.  Next we find the distance between
$\mathbf{p}(D_i)$ and our precomputed baseline, as well as the distance
between $\mathbf{p}(D_i)$ and $\mathbf{p}(D_0)$ to monitor the drift rate.

\figurename~\ref{fig-wander-background} shows the results for four 
topics as starting points.  Because all the starting points in each
group were taken from a specific topic, the topic histogram at
distance 0 is quite dissimilar from the background.  Even 20 hops seem
inadequate to bring the distribution closer to the background.  But
this is not because the walks stay perfectly on-topic.  They clearly
do start drifting away, but not too badly within the first 5--10 hops.
Thus it is clear that a sampling-type walk is critical to topic
convergence.  The rate of drift away from the starting topic also
varies from topic to topic: `Soccer' seems to be very drift-prone
whereas `Photography' drifts much less.  Such prior estimates would be
valuable to a focused crawler, and explain in part why focused
crawling along forward links is already quite successful
\cite{ChakrabartiVD1999focus1}.  We also confirm our intuition that
our estimate of drift w.r.t.\ broad topics seems generally lower than
what Davison and Menczer have characterized in terms of cosine
similarity with the starting point.

\subsection{Implications}

Two families of popularity-based ranking of Web pages have been very
successful in recent years.  Google (\url{http://google.com}) uses as
a subroutine the PageRank algorithm
\cite{BrinP1998anatomy,PageBMW1998pagerank}, which we have already
reviewed in \S\ref{sec-intro-pagerank}.  Kleinberg proposed the
HITS algorithm \cite{Kleinberg1998hits} which has many variants.

Unlike PageRank, HITS does not analyze the whole Web graph, but
collects a subgraph $G_q=(V_q,E_q)$ of the Web graph $G$ in response
to a specific query $q$.  It uses a keyword search engine to collect a
\emph{root~set} $R_q$, includes $R_q$ in $G_q$ and then further
includes in $G_q$ any node linked from or linking to some node in $R_q$
(a radius-one link expansion).  In matrix terms, we can reuse $E_q$ to
denote the node adjacency matrix: $E_q(i,j)=1$ if $i$ links to $j$,
and 0 otherwise.  HITS assigns \emph{two} scores $a(v)$ and $h(v)$
with each node, reflecting its \emph{authority} and its \emph{hubness}
(the property of compiling a number of links to good authorities), and
solves the following mutually recursive equations iteratively, scaling
$\mathbf{a}$ and $\mathbf{h}$ to unit norm every step:
\begin{eqnarray}
\mathbf{a} \leftarrow E_q^T\mathbf{h} &\mathrm{and}&
\mathbf{h} \leftarrow E_q\mathbf{a}.
\end{eqnarray}

It has been known that the radius-one expansion improves recall; good
hubs and authorities which do not have keyword matches with the query
may be drawn into $G_q$ this way.  Some irrelevant pages will be
included as well, but our experiments confirm that
at radius one the loss of precision is not devastating.
Davison and Menczer
\cite{Davison2000locality,Menczer2001locality} have proposed that
topical locality is what saves $G_q$ from drifting too much, but they
did not use a reference set of topics to make that judgment.

Unlike HITS, PageRank is a \emph{global} computation on the Web graph,
which means it assigns a query-independent prestige score
to each node.  This is faster than collecting and analyzing
query-specific graphs, but researchers have hinted that the
query-specific subgraphs should lead to more faithful scores of
authority \cite{Kleinberg1998hits}.

PageRank has a random jump parameter $d$, which is empirically set
to about 0.15--0.2.  This means that typically, every sixth to eighth
step, the random surfer jumps rather than walks to an out-neighbor,
i.e., the surfer traces a path of typical length 6--8 before
abandoning the path and starting afresh.  The \emph{key observation}
is this: if topic drift is small on such short directed random paths,
as \figurename~\ref{fig-wander-background} seems to indicate, the
global nature of the PageRank computation does not hurt, because
endorsement to any node (apart from jumping to it uar, which treats
all nodes equally) comes from a small neighborhood which is topically
homogeneous anyway!

All this may explain why PageRank is an acceptable measure of prestige
w.r.t.\ \emph{any} query, in spite of being a global measure.
Confirming this intuition will not be easy.  Perhaps one can build
synthetic graphs, or extract large graphs from the Web which span
multiple topic communities, and tweak the jump probability $d$ so that
the average interval between jumps ranges from much less than directed
mixing radius to much more that mixing radius, and see if the largest
components of the PageRank vector leaves a variety of topic
communities to concentrate in some dominant communities featuring
``universal authorities'' such as
\url{http://www.yahoo.com/}, \url{http://www.netscape.com/},
\url{http://www.microsoft.com/windows/ie/}, or
\url{http://www.adobe.com/prodindex/acrobat/readstep.html}.


\section{Relations between topics\label{sec-cite}}

According to the ordinary graph model for Web pages, a link or edge
connects two nodes which have only graph-theoretic properties such as
degree.  Given our interest in the \emph{content} of pages, it is
natural to extend our view of a link as connecting two \emph{topics}.
The topic of a potential link target page $v$ is clearly the single
most important reason why the author of another page $u$ may be
motivated to link from $u$ to $v$.

This view is clearly missing from  preferential attachment theory,
where the author picks targets with probability proportional to their
current indegree, \emph{regardless} of content or topic.  It would be
very interesting to see a general, more realistic theory that folds
some form topic affinity with preferential attachment and matches our
observations.

We model topic affinity using a \emph{topic citation matrix},
which is constructed using soft-counting as follows:
\begin{enumerate}
\item
Suppose there are $N$ topics at the desired level of detail.
Initialize a $N\times N$ topic citation matrix $C$ with all zeroes.
\item
Repeat the following steps as long as the estimate $C$ `improves'.
(E.g., if $\ell$ links have been sampled, we want $C/\ell$ to
converge.)
\begin{enumerate}
	\item
	Sample a page $u$ nearly uar from the Web using the \textsc{Sampling}
	walk and the virtual hop technique.
	\item
	Sample an outlink $v$ of $u$ uar.
	\item
	Fetch and classify $u$ and $v$, getting class probability
	row vectors $\mathbf{p}(u)$ and $\mathbf{p}(v)$.
	\item
	For every position $(i,j)$ in $C$, increase $C(i,j)$ by
	$p_i(u)\,p_j(v)$.  In matrix terms, assign
	$C \leftarrow C + \mathbf{p}(u)^T \mathbf{p}(v)$.
\end{enumerate}
\end{enumerate}
The hard-counting counterpart would classify $u$ and $v$ to their
best-matching classes $\gamma_u$ and $\gamma_v$, and increment
$C(\gamma_u,\gamma_v)$ by one.  If rows of $C$ are scaled to~1, the
entry $C(i,j)$ gives the empirical probability that a random outlink
from a page about topic $i$ will link to a page about topic $j$.

\begin{figure}[hb]
\begin{center} \tabcolsep0pt \footnotesize
\begin{tabular}{cc}
\iftth
\tthfig{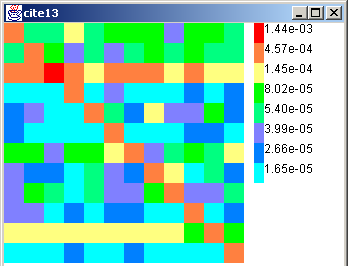} &
\tthfig{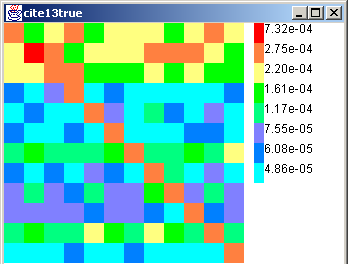}
\else
\tthfig{\includegraphics*[width=.5\hsize]{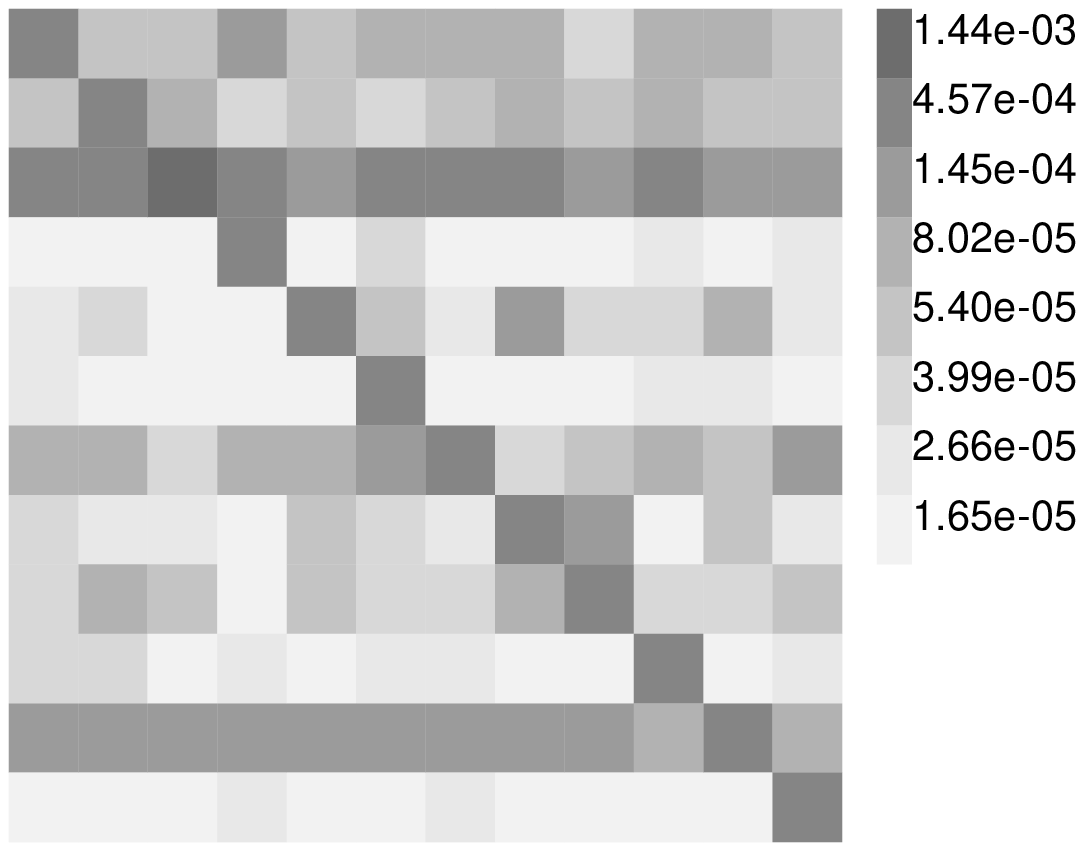}} &
\tthfig{\includegraphics*[width=.5\hsize]{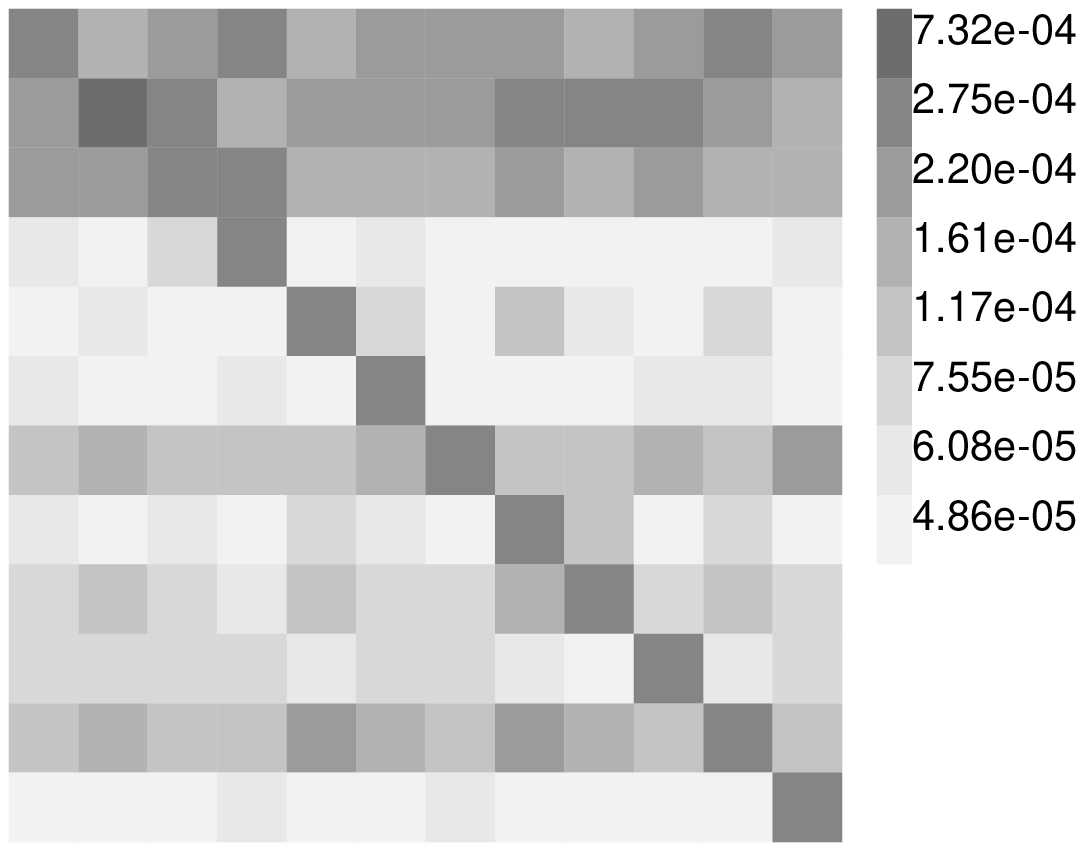}}
\fi
\\ (a)~Raw citation & (b)~Confusion-adjusted
\end{tabular}
\end{center}
\caption{(a)~Citation matrix for the 12 top-level topics.
The source topic runs horizontally to the right, the target
topic runs vertically down.  (b)~Confusion-adjusted
citation matrix, see text later.}
\label{fig-cite12n}
\end{figure}

\subsection{Experimental results}

We experimented with our 482-leaf taxonomy at two levels of detail:
the top level with 12 topics and the third level with 191 topics.
This is partly because data at the $482\times482$ level was very
sparse.  In \figurename~\ref{fig-cite12n}(a), we show the 12-class
top-level citation matrix.  Dark colors (in the HTML version, hot
colors) show higher probabilities.  The diagonal is clearly dominant,
which means that there is a great deal of self-citation among topics.
This natural social phenomenon explains the success of systems like
HITS and focused crawlers.  It is also worthwhile to note that the
matrix is markedly asymmetric, meaning that communities do not 
reciprocate in cross-linking behavior.

Apart from the prominent diagonal, there are two horizontal bands
corresponding to \path{/Computers} and \path{/Society}, which means
that pages relevant to a large variety of topics link to pages about
these two topics and their subtopics.  Given that these are the two
most dominant top-level topics on the Web
(\figurename~\ref{fig-back12u}), it is conceivable that preferential
attachment will lead to such behavior.

\begin{figure}[ht]
\begin{center} \tabcolsep0pt \footnotesize
\begin{tabular}{cc}
\iftth
\tthfig{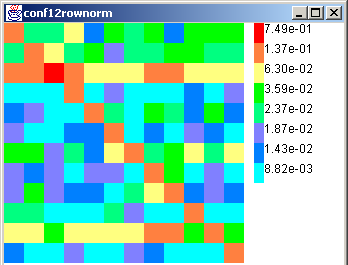} &
\tthfig{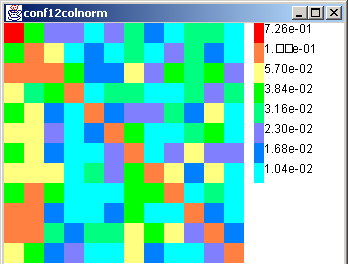}
\else
\tthfig{\includegraphics*[width=.5\hsize]{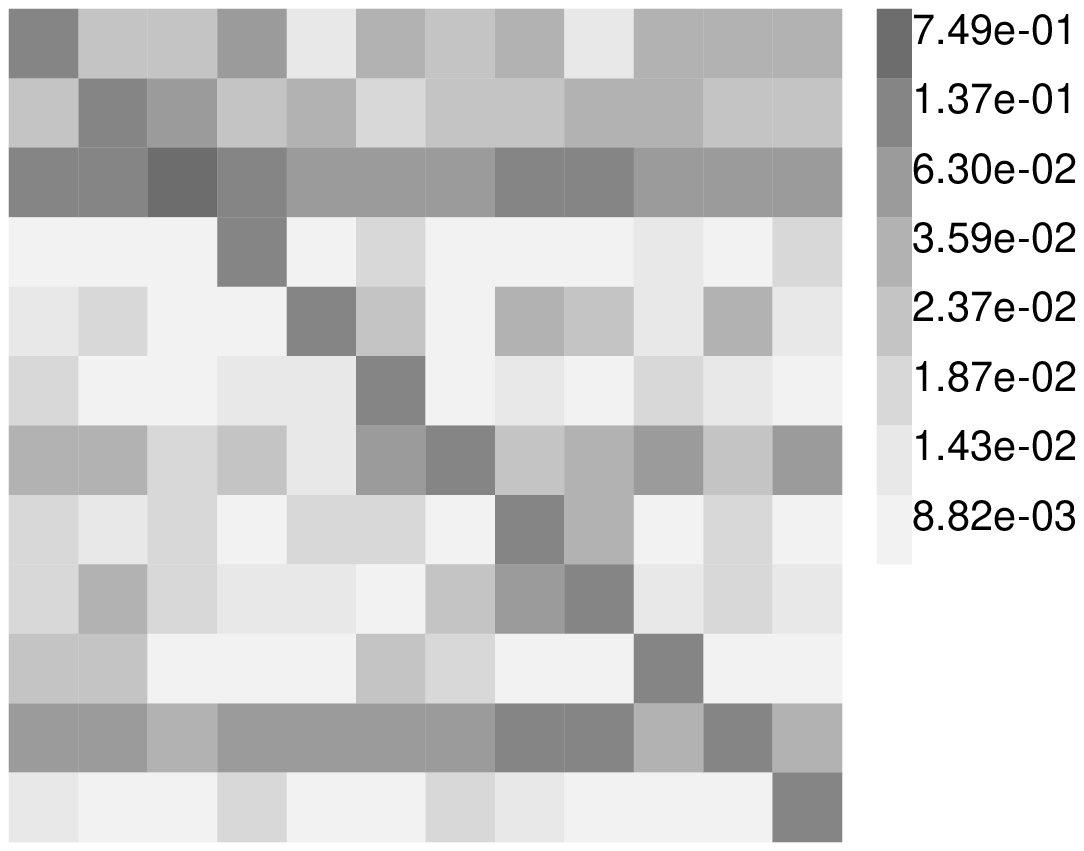}} &
\tthfig{\includegraphics*[width=.5\hsize]{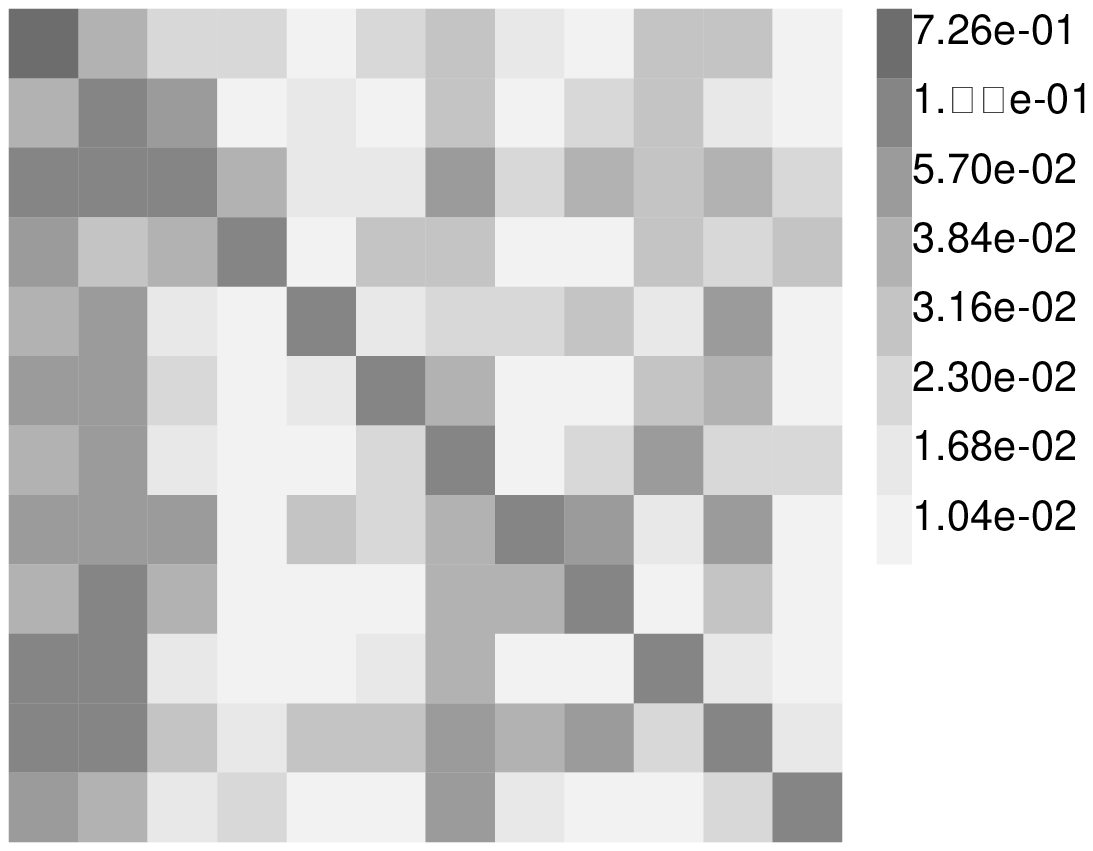}}
\fi
\\ True-normalized & Guess-normalized
\end{tabular}
\end{center}
\caption{The confusion matrix between topics at the top level.
``True-normalized'' means that the number of test documents from each
class is scaled to~1.  ``Guess-normalized'' means that the number of
documents tagged with each class by the classifier is scaled to~1.}
\label{fig-conf12norm}
\end{figure}

However, these inferences may be specious if it turns out that our
classifier is biased in favor of \path{/Computers} and \path{/Society}
when it guesses the class of test documents.  To avoid this problem,
we use a held-out labeled data set from DMoz to calibrate the
classifier.  The result is a \emph{confusion matrix} $E$ where
$E(i,j)$ is the number of documents in class $i$ labeled with class
$j$ by the classifier.  Depending on the application, we can scale
either the row (true class) or the column (guessed class) of $E$ to~1.
In either case, an ideal classifier's confusion matrix will be the
identity matrix.  We show both scaled versions $E_t$ and $E_g$ in
\figurename~\ref{fig-conf12norm}.  Although the color scales were 
designed for maximum contrast, we note that the diagonal elements are
generally large, hinting that the classifier is doing well.

The entries corresponding to guessed class $j$ in the
``guess-normalized'' version $E_g$ may be regarded as the empirical
$\Pr(\mathrm{true}=i|\mathrm{guess}=j)$ for all~$i$.
Information from $E_g$ can be folded into the soft counting process as
follows.  After we fetch and classify $u$ and $v$, getting class
probability row vectors $\mathbf{p}(u)$ and $\mathbf{p}(v)$, we find
the corrected probability vector $\mathbf{p^*}(u)$ (similarly,
$\mathbf{p^*}(v)$) by computing the corrected probability that $u$
`truly' belongs to class $i$ as
\begin{eqnarray}
\mathbf{p^*}_i(u) &=& \sum_j E_g(i,j) \mathbf{p}_j(u),
\end{eqnarray}
under certain technical assumptions on the sampling process that we
omit mentioning for lack of space.  The citation matrices can be
corrected accordingly.  In fact, this technique can/should also be
used to correct the background distribution
(\figurename~\ref{fig-back12u}).
Although the sharp notches at \path{/Computers} and \path{/Society}
get somewhat subdued, our broad observations remain valid.

Returning to the citation matrix, we present the corrected citation
matrix in \figurename~\ref{fig-cite12n}(b).  Generally speaking, the
correction smears out the probability mass slightly, but the corrected
data continues to show a higher than average rate of linkage to
documents about \path{/Computers} and \path{/Society}.

\begin{figure}[ht]
\begin{center}
\iftth
\tthfig{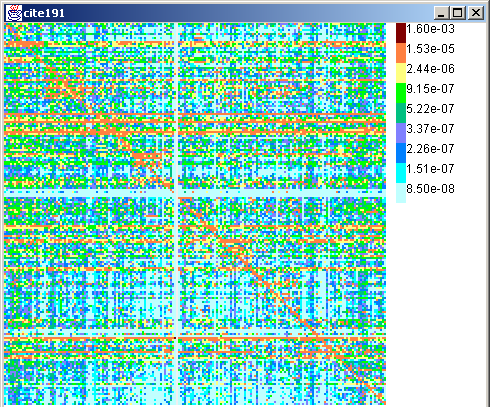}
\else
\tthfig{\includegraphics*[width=\hsize]{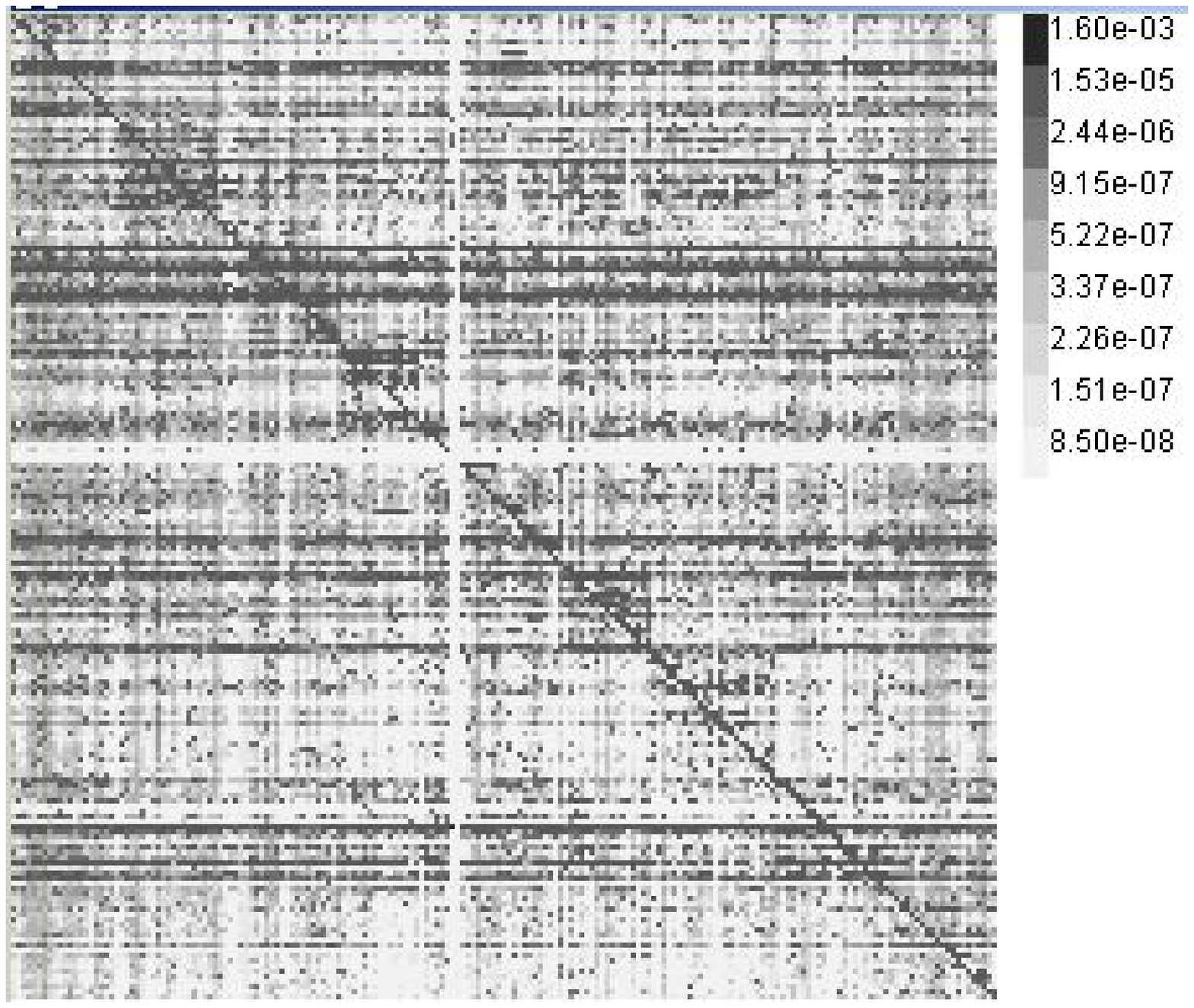}}
\fi
\caption{Citation matrix at the third level with 191 topics.}
\label{fig-cite191n}
\end{center}
\end{figure}

We can now extend the experiments to more detailed levels in the
taxonomy.  \figurename~\ref{fig-cite191n} shows the citation matrix at
the third level of our DMoz sample, with 191 topics.  This diagram is
a drill-down into the earlier 12-class citation diagram, and we see a
telltale block-diagonals and other block-structure in the larger
matrix, which align with the broad 12-topic boundaries.  (They are not
all equally wide, because the top-level topics have different numbers
of descendants.)

The diagonal remains dominant, which is good news for topic
communities and focused crawling.  Finer horizontal lines emerge,
showing us the most popular subtopics under the popular broad topics.
Zooming down into \path{/Arts}, we find the most prominent bands are
at \path{/Arts/Music}, \path{/Arts/Literature} and
\path{/Arts/Movies}.  Within \path{/Computers}, we find a deep, sharp 
line about a fourth of the way down, at \path{/Computers/DataFormats}.
This is partly an artifact of badly written HTML which confounded our
\path{libxml2} HTML parser, making HTML tags part of the 
classified text.  Other, more meaningful target bands are found at
\path{/Computers/Security}, \path{/Recreation/Outdoors}, and
\path{/Society/Issues}.
We found many meaningful isolated hot-spots, such as from
\path{/Arts/Music} to \path{/Shopping/Music} and 
\path{/Shopping/Entertainment/Recordings}.

\begin{figure}[ht]
\begin{center}
\iftth
\tthfig{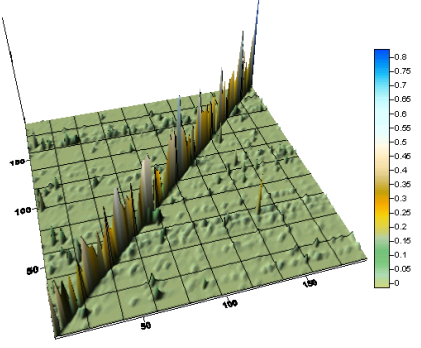}
\else
\tthfig{\includegraphics*[width=\hsize]{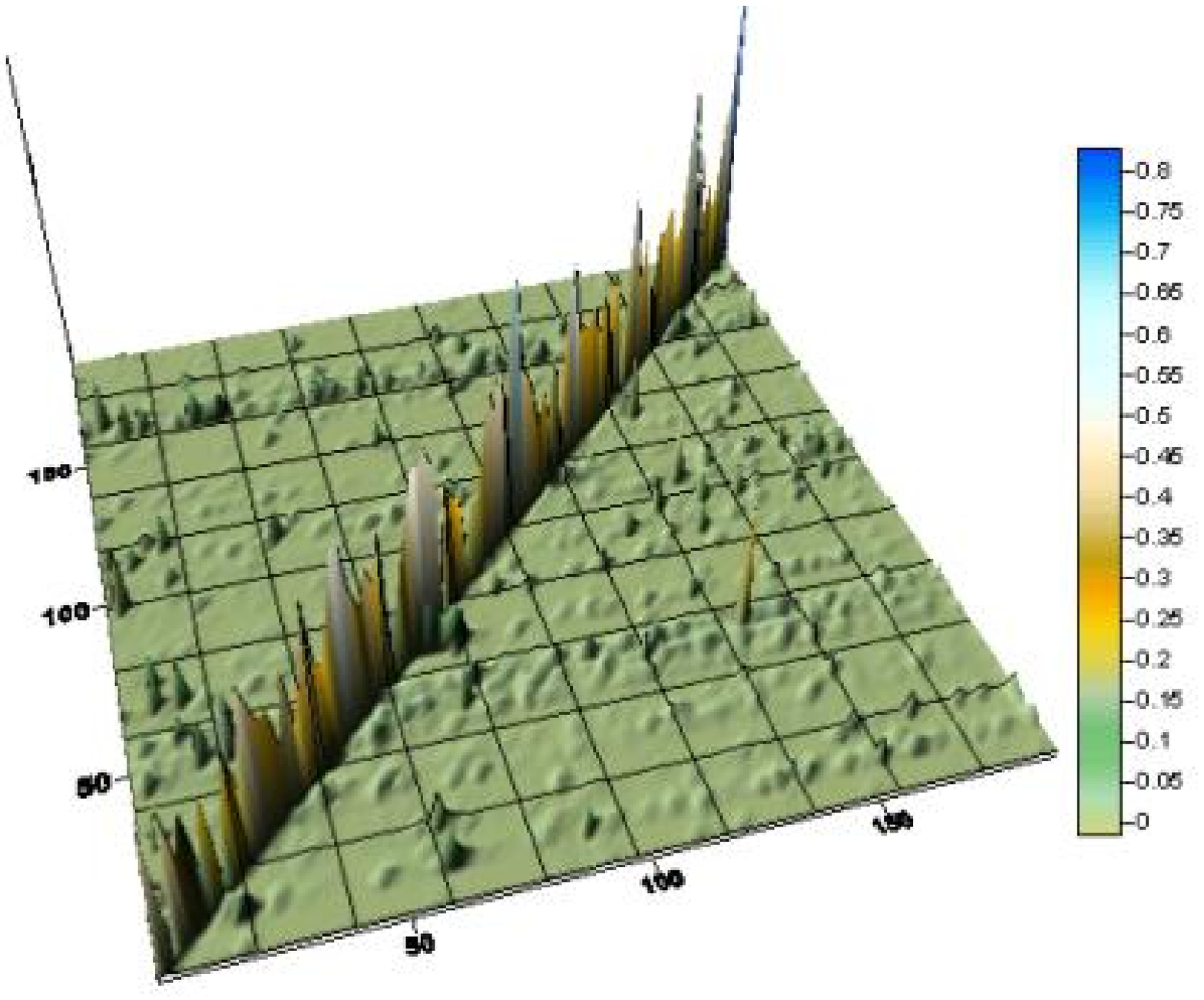}}
\fi
\caption{Contour plot of the 191-topic citation matrix.}
\label{fig-cite191ns}
\end{center}
\end{figure}

Because the `height' at various pixels is not very clear from a 2d
rendering, we also show a 3d contour-plot of the $191\times191$
citation matrix in \figurename~\ref{fig-cite191ns}.  Although the
excessive smoothing and the
slope introduced between adjacent pixels is artificial, the
contour-plot does reveal how strongly self-citing the topics are, even
at the third level of DMoz.


\subsection{Implications and applications}

The topic citation matrix,
similar in spirit to a spectrogram,
is fascinating in its own right in showing
the strongest (single-link) inter-topic connections on the Web, but it
also has a variety of practical applications.  We touch upon a few here.

\paragraph{Improved hypertext classification:}
Standard Bayesian text classifiers build a class-conditional estimate
of document probability, $\Pr(d|c)$, in terms of the textual tokens
appearing in $d$.  Pages on the Web are not isolated entities, and the
(estimated) topics of the neighbors $N(u)$ of page $u$ may lend
valuable evidence to estimate the topic of $u$
\cite{ChakrabartiDI1998hyperclass,JinD2001textlink}.
Thus we need to estimate a joint
distribution for $\Pr\bigl(c(u),c(N(u))\bigr)$, 
which is a direct application of
the topic citation matrix.

\paragraph{Enhanced focused crawling:}
A focused crawler is given a topic taxonomy, a trained topic
classifier, and sample URLs from a specific topic.  The goal is to
augment this set of relevant URLs while crawling as few irrelevant
pages as possible.  Currently, focused crawlers use the following
policies:
\begin{itemize}
\item
If page $u$ is relevant, an outlink $v$ is also likely to be relevant
\cite{ChakrabartiVD1999focus1}.
\item
Given the text of a page $u$, we can estimate the link distance to a
relevant goal node \cite{DiligentiCLGG2001context}.
\item
Given features from near outlink $(u,v)$ on page $u$, estimate the
`reward' that may be accrued by following the link
\cite{ChakrabartiPS2002focus,RennieM1999reinforce}.
\end{itemize}
In addition to learning the above patterns, it may be very valuable
for a focused crawler to have access to the topic linkage matrix.
Consider the topic \path{/Shopping/Consumer_Electronics}.
Examples of this topic are often competing sites which do not wish
to link to each other.  But they sell cameras, and are therefore
cited by pages about \path{/Arts/Photography}, many of which are
not owned by businesses and are heavily linked up.  Such patterns
can help the focused crawler traverse between relevant islands on the
Web.

\paragraph{Reorganizing topic directories:}
We had to discard the \path{/Regional} and \path{/News}
subtrees, not only because of severe classification
error, but also because of overwhelming interlinkage between these
and other topics.  \path{/News} is almost always \emph{about} some other
topic, such as \path{/Sports} or \path{/Science}.  
The structure within the \path{/Regional} or even the
\path{/Shopping} subtree mirrors many broad topics outside.
E.g., mountain biking fits under recreation
and sports, but to buy biking gear, one must move up and then descend
down the \path{/Shopping} path.  A tree is a rather inadequate
structure to connect areas of human endeavor and thought, and such
artificial structures have compelled Yahoo! to include a number of
``soft links'' connecting arbitrary points in their topic `tree'.

The limitations of a tree representation are responsible for many
long-range off-diagonal elements in the topic citation matrix, and
arguably makes directory browsing less intuitive.  It may also make
focused crawling and automatic cataloging more difficult.  We claim
that such phenomena warrant careful consideration of \emph{taxonomy
inversion} and better metadata annotation.
E.g., let commercial interests related to biking be
contained in the subtree dedicated to biking, rather than collect
diverse commercial interests together.  We can envisage advanced user
interfaces through which taxonomy editors can point and click on a
topic citation map or a confusion matrix to reorganize and improve the
design of a taxonomy.

\section{Concluding remarks}

The geography of the Web, delineating communities and their boundaries
in a state of continual flux, is a fascinating source of data for
social network analysis (see, e.g.,
\url{http://www.cybergeography.org/}).  
In this paper we have initiated
an exploration of the terrain of broad topics in the Web graph, and
characterized some important notions of topical locality on the Web.
Specifically, we have shown how to estimate the background
distribution of broad topics on the Web, how pages relevant to these
topics cite each other, and how soon a random path starting from a
given topic `loses' itself into the background distribution.

We believe this work barely scratches the surface w.r.t.\ a new,
content-rich characterization of the Web, and opens up many questions,
some of which we list below.

\paragraph{PageRank jump parameter:}
How should one set the jump probability in PageRank?  Is it useful to
set topic-specific jump probabilities?  Does an understanding of
mixing radius help us set better jump probabilities?  Is there a
useful middle ground between PageRank's precomputed scores and HITS's
runtime graph collection?

\paragraph{Topical stability of distillation algorithms:}
How can we propose models of HITS and stochastic variants that are
content-cognizant?  Can content-guided random walks be used to
\emph{define} what a focused crawler should visit and/or collect?  Can
we validate this definition (or proposal) on synthetic graphs?  Can
such a theory, coupled with our measurements of topic linkage, predict
and help avoid topic drift in distillation algorithms?



\paragraph{Better crawling algorithms:}
Given that we can measure mixing distances and inter-topic linkage,
can we develop smarter federations of crawlers in which each
concentrates on a collection of tightly knit topics?  Can this lead to
better and fresher coverage of small communities?  Can we exploit the
fact that degrees follow power laws both globally and locally within
topic contexts to derive better, less topic-biased samples of URLs
from the Web?


\paragraph{Acknowledgments:}
Thanks to Alice Zheng and David Lewis for helpful discussions, to Kim
Patch for bringing Menczer's recent measurements to our notice, to
Gary W.\ Flake for providing the NEC crawl data, and for his help with
our code.  We also thank the referees for helpful comments.

\iftth\else \begingroup \footnotesize \fi
\bibliographystyle{abbrv}
\bibliography{c1,c2,c3,c4,c5,c6,c7,c8,c15,c20}
\iftth\else \endgroup \fi

\end{document}